\documentclass[aps,twocolumn,showpacs,nofootinbib,showkeys,superscriptaddress,preprintnumbers,longbibliography,10pt]{revtex4-2}
\usepackage[utf8]{inputenc}

\usepackage{latexsym}
\usepackage{epsfig}
\usepackage{graphicx,subfigure}
\usepackage[normalem]{ulem}
\usepackage{color}
\usepackage{xcolor}
\usepackage{multirow}
\usepackage{hyperref}
\usepackage{numprint}
\usepackage{eucal}
\usepackage{amsmath}
\usepackage{amssymb}
\usepackage{amsmath,amssymb} 
\usepackage{wasysym}

\newcommand*{\mathabxbfamily}{\fontencoding{U}\fontfamily{mathb}\selectfont}
\DeclareFontFamily{U}{mathb}{\hyphenchar\font45}
\DeclareFontShape{U}{mathb}{m}{n}{
      <5> <6> <7> <8> <9> <10> gen * mathb
      <10.95> mathb10 <12> <14.4> <17.28> <20.74> <24.88> mathb12
      }{}
\newcommand*{\Sun}{{\text{\mathabxbfamily\char"40}}}

\newcommand*{\Earth}{{\text{\mathabxbfamily\char"43}}}

\definecolor{blue}{rgb}{0,120,250}

\usepackage{xcolor}

\newcommand{\ssout}[1]{}
\begin{document}

\def\ga{\mathrel{\raise.3ex\hbox{$>$\kern-.75em\lower1ex\hbox{$\sim$}}}}
\def\la{\mathrel{\raise.3ex\hbox{$<$\kern-.75em\lower1ex\hbox{$\sim$}}}}

\def\be{\begin{equation}}
\def\ee{\end{equation}}
\def\bea{\begin{eqnarray}}
\def\eea{\end{eqnarray}}

\def\betap{\tilde\beta}
\def\del{\delta_{\rm PBH}^{\rm local}}
\def\Msun{M_\odot}

\newcommand{\dd}{\mathrm{d}} 
\newcommand{\Mpl}{M_P} 
\newcommand{\mpl}{m_\mathrm{pl}} 

\newcommand{\CHECK}[1]{{\color{red}~\textsf{#1}}}

\title{Probing Primordial Black Holes and Dark Matter Clumps in the Solar System with Gravimeter and GNSS Networks.}

\author{Michal Cuadrat-Grzybowski}
\affiliation{Space Engineering Department, Delft University of Technology, Kluyverweg 1
2629 HS Delft, (the) Netherlands}

\author{S\'ebastien Clesse}
\affiliation{Service de Physique Th\'eorique, Universit\'e Libre de Bruxelles (ULB), Boulevard du Triomphe, CP225, B-1050 Brussels, Belgium}

\author{Pascale Defraigne}
\affiliation{Royal Observatory of Belgium, Avenue Circulaire, 3, 1180 Uccle, Belgium}

\author{Michel Van Camp}
\affiliation{Royal Belgian Institute of Natural Sciences, Rue Vautier 29, 1000 Brussels, Belgium}

\author{Bruno Bertrand}
\affiliation{Royal Observatory of Belgium, Avenue Circulaire, 3, 1180 Uccle, Belgium}

\begin{abstract} 
%
We show that Global Navigation Satellite Systems (GNSS) and gravimeters on Earth and in space can potentially offer the most accurate direct measurement of local density of near-Earth asteroid-mass Primordial Black Holes (PBHs) and Dark Matter (DM) clumps in the solar system by means of gravitational influence.
Using semi-analytical methods and Monte Carlo simulation, this paper revisits the analysis of the trajectories of DM clumps in the solar system, including both captured objects and hyperbolic trajectories. A link is thus made between the frequency and distance of Earth overflights for a given mass flux, and a direct measure of dark matter clump density in the solar system.
We then model the signature of a close flyby of a DM object on orbital data from GNSS satellites and gravity measurements from gravimeters. We thus obtain a first assessment of the single probe sensitivity.
It paves the way for an exhaustive statistical analysis of 28 years of gravimeters and GNSS data to obtain observational constraints on the density of the PBHs and DM clumps within the solar system, for the mass range $[10^8-10^{17}]$ kg. In addition, our methodology offers a possibility of direct detection in cases where DM clumps are endowed with an additional long-range clump-matter fifth-force beyond gravity.

\end{abstract}

\maketitle

\section{Introduction}
\label{sec:Introduction}

The nature of Dark Matter (DM), representing about $27\%$ of the total mass-energy density of the Universe~\cite{Planck:2018vyg}, is still unknown and has remained since more than fifty years one of the deepest mysteries in Physics.  Whereas there are multiple indirect probes of the existence of DM on galactic, extra-galactic and cosmological scales, there is currently no observational evidence coming from smaller scales.  
It is therefore plausible that DM clusters into compact sub-galactic structures, possibly fragmenting into even smaller parts, referred here as DM clumps. The particle nature of DM is also still a matter of debate and there are various theoretical scenarios in which all the DM, or a significant fraction of it, is made of massive compact dark objects. These scenarios include e.g. primordial black holes (PBHs)~\cite{Hawking:1971ei,Carr:1974nx}, dark quark nuggets~\cite{Witten:1984rs}, dark blobs or other composite states~\cite{Wise:2014jva,Grabowska:2018lnd}, strangelets~\cite{Witten:1984rs}, nontopological solitons~\cite{Kusenko:2001vu}, mirror DM \cite{Blinnikov:1983gh,Khlopov:2013ava}, axion or scalar miniclusters~\cite{Hogan:1988,Enander:2017ogx}, axion~\cite{Seidel:1993zk,Braaten:2018nag} or boson stars~\cite{Colpi:1986ye,Eby:2015hsq}...

This work focuses on 
DM clumps, compact objects or PBHs with a mass between 10$^{11}$ to 10$^{20}$ kg. This mass range is especially relevant because microlensing of stars becomes ineffective in detecting compact DM objects below {10$^{20}$ kg~\cite{Montero-Camacho:2019jte}.  
%
PBHs that are by definition black holes formed in the early  Universe, are further constrained due to their expected evaporation through Hawking radiation.  Below $10^{11}$ kg, they evaporate in a shorter time than the age of the Universe and so are not expected to survive today.  There is no constraint between $10^{14}$ and $10^{20}$ kg.  Between $10^{11}$ and $10^{14}$ kg, there are some constraints on their abundance from their partial evaporation, but these are subject to large uncertainties~\cite{Auffinger:2022dic} and rely on the still speculative evaporation process.
Altogether, it is possible that asteroid-mass compact DM objects or PBHs~\cite{Montero-Camacho:2019jte} constitute a significant fraction or even the totality of the DM. }


As those DM clumps or compact objects travel in our galaxy, they eventually pass through the Solar System, itself in motion around the Galactic center towards the Cygnus constellation.  This opens the possibility of a close approach of the Earth~\cite{Grabowska:2018lnd, Foot:2001cq, Seto:2004zu,*Adams:2004pk, Derevianko:2013oaa,Hall:2016usm, Visinelli:2018wza,Banerjee:2019epw,Baum:2022duc}, provided the DM clumps are sufficiently stable and dense not to be tidally disrupted as they pass through the inner solar system. 
In addition, some of them could be trapped by the Sun's gravitational potential, thereby increasing the local DM density~\cite{Pitjev:2013sfa}. For PBHs, it was even proposed that they could explain the suspected Planet~9~\cite{Scholtz:2019csj}.  PBHs in the Oort clouds could be detected if they accrete matter~\cite{Siraj2020}, or using continuous gravitational-wave searches~\cite{Miller2020} if there are PBH binaries.


To the best of our knowledge, there are currently no direct experimental measurements of the density of DM clumps or PBHs in the solar system, for masses ranging from 10$^{11}$ to 10$^{20}$ kg. The only constraints that exist are based on extrapolations of local dark matter density measurements in the galactic region where the solar system is located. These indirect constraints are obtained by including the effects of accretion, capture and ejection by solar system bodies in the case of dark matter particles such as Weakly Interacting Massive Particles (WIMPS) \cite{Edsjo2010, Iorio2010, Khriplovich2009, Xu2008}.
In addition, the density of DM particles like WIMPS in the solar system could be probed using the signals from space probes or the precise ephemeris of planets \cite{Pitjev:2013sfa}. However, it will be discussed in section \ref{sec:ephemeris} that such constraints are not transferable to the case of DM clumps or PBHs of asteroidal mass. Finally, gravitational-wave (GW) detectors can also be used to observe 
such dark objects and constrain their abundance \cite{Seto:2004zu,*Adams:2004pk}. Indeed, close overflights of DM clumps or PBHs are supposed to exert an acceleration on the test masses of a GW detector. For example, the LIGO/Virgo detectors is suitable for masses below 10$^5$ kg only~\cite{Hall:2016usm} given that the detector sensitivity is optimal at frequencies between $10$ and $10^4$ Hz. In the future, the Laser Interferometer in Space Antenna (LISA) operating at much lower frequencies will be sensitive to objects with a mass ranging from $10^8$ to $10^{15}$ kg due to the very long baselines~\cite{Hall:2016usm,Baum:2022duc}. However, the expected rate of discoverable signals remains limited to $3\times 10^{-3}$ for a purely gravitational clump-matter interaction, well beyond the detector's lifetime. Finally, the recent analysis by \cite{Tran:2023jci} proposes to use the measurement of the distance between the Earth and the other planets of the inner solar system to detect PBHs flybys at a few AU. However, the sensitivity is limited to PBHs with a mass greater than $10^{16}$~kg.

The aim of this study is to propose a methodology for setting the first constraints on the density of DM clumps in the solar system by direct observation.
For those DM clumps or objects passing sufficiently near the Earth,
tiny changes in the gravitational field could be detected by global navigation satellite systems (GNSS), and/or gravimeters on Earth and in space. For GNSS, like the US GPS, the EU Galileo, the Chinese BeiDou or the Russian Glonass constellations, any change in the gravitational potential should affect satellite orbits. The precise atomic clocks onboard GNSS satellites along with GNSS signal carrier phase analysis allows an accurate determination of these orbits, reaching a precision at the centimeter level. Given that up to 28 years of orbit data are publicly available, one has at disposal many years of data from an Earth-size DM detector that have not been exploited so far. For superconducting gravimeters, the world record of the longest measurement of variations of a local gravitational field vy one instrument is detained by the Belgian superconducting gravimeter in Membach~\cite{VanCamp:2017}, operated by the Royal Observatory of Belgium since 1995. The level of precision is of order $10^{-11} g$ within 1 minute (where $g\simeq 9.81 {\rm m/s^2}$ is the Earth gravitational acceleration).  
The goal of the present paper is to study for the first time how to use and combine GNSS and gravimeter data in order to post-detect near-Earth DM clumps or compact objects like PBHs, or to set limits on their abundance in the solar system.

Our work takes place in line with a recent tendency to test fundamental Physics with geodetic tools such as GNSS \cite{Derevianko:2013oaa, Roberts:2017hla, Stadnik:2020bfk, Delva:2015kta, Delva:2018ilu, Herrmann:2018rva} and superconducting gravimeters \cite{Horowitz2019,Hu2020,Namigata2022}, using data mining over the long-term time series. For example, Refs. \cite{Delva:2015kta, Delva:2018ilu, Herrmann:2018rva} seized the opportunity to transform a failure into a success: because of a technical problem, a launcher brought in 2014 two Galileo satellites on a wrong orbit with a large eccentricity. The resulting modulation of the gravitational redshift of the onboard atomic clocks allowed the redshift determination with high accuracy, thus making a precise test of General Relativity. Other works \cite{Derevianko:2013oaa, Roberts:2017hla, Stadnik:2020bfk} tested DM models made of macroscopic topological defects based on light scalar fields coupling quadratically with the Standard-model fields. The resulting apparent spatial variation of the fundamental constants should affect the GNSS atomic clock frequency. Hence, strong constraints have been brought on models either through a direct transient signature or non-transient signatures due to the back-action of the Earth's mass on the scalar field.
Recently, gravimeters from the Black Forest and Northern Benin were also used in~\cite{Horowitz2019,Namigata2022} to set limits on light PBHs orbiting around or inside the Earth but without being able to set a significant limit on their cosmological abundance.  

Here, we propose to jointly exploit and cross-correlate gravimetry and GNSS data to track anomalies in the gravitational potential of space origin, e.g. induced by DM clumps or PBHs.  For GNSS in particular, this is the first study that calculates their direct gravitational influence on the satellite orbits. For gravimeters, we investigate the case of transient signatures whereas previous work only focused on the periodic signatures of PBHs captured by the Earth, which we find to be much less likely than near-Earth passages.   

Our analysis also goes beyond simple approximations by numerically integrating trajectories in the solar system and by performing Monte-Carlo simulations of a realistic DM clumps/objects population, which allows us to fully take into account the gravitational focusing effect of the Sun and to compute more precisely the rate of three-body captures induced by the different planets of the solar system.  The final product of our analysis relates the GNSS and gravimeter sensitivity to the expected number of detectable events and the resulting limit on the cosmological abundance of DM clumps/objects.

The paper is organized as follows. In Section~\ref{sec:theory_and_simple_estimates} we introduce the theoretical scenarios leading to asteroid-mass DM clumps or compact objects, such as PBHs, as well as the relevant limits on their abundance.  A quick first estimate of the expected number of events and the possible limits from GNSS and gravimeters is realized, using simplifying assumptions.  In Section~\ref{sec:Trajectories} we discuss the hyperbolic trajectories of unbound objects under the Sun gravitational influence. The event rate of Earth close fly-by is then estimated. The rate of capture by solar system planets is analysed in Section~\ref{sec:DM_capture}. The expected signatures in gravimeter and GNSS orbit data are modeled in Section~\ref{sec:expected_signatures}, which can be compared to the sensitivities estimated in Section~\ref{sec:theory_and_simple_estimates}.  
In Section~\ref{sec:Sensitivity}, a preliminary analysis based on superconducting gravimeter data series and Galileo orbital solutions will provide a first assessment of the ‘one-probe’, i.e. one satellite or one gravimeter, sensitivity. Our conclusion and the perspectives of our work are presented in Section~\ref{sec:Conclusion}.




\section{Theoretical Models and simple estimates} \label{sec:theory_and_simple_estimates}

\subsection{Dark Matter Clumps}
Some models predicts that the DM halo clusters into compact sub-galactic structures, possibly fragmenting into even smaller parts.
If DM is made of particles, there is no theoretical lower limit on the mass of DM structures, because there is no lower bound on the self-interaction strength between dark matter particles nor between dark matter and standard model particles, and because the possible mass of DM particles is theoretically unconstrained.  Limits on the mass of DM particles can nevertheless appear in fixed theoretical frameworks, e.g. a limit of $10^{-18}$ eV was recently claimed assuming that DM particles are produced after inflation through a process with a finite correlation length~\cite{Amin:2022nlh}.  A similar bound of $3 \times 10^{-19}$ eV was obtained for fuzzy dark matter, coming from the dynamical heating of stars in ultra-faint dwarf galaxies~\cite{Dalal:2022rmp}.  The DM particle mass $m_{\rm DM}$ is relevant for dark matter clumps because it enters in the associated De Broglie wavelength, $\lambda = h / (m_{\rm DM} v_{\rm DM}) $ for non-relativistic particles, where $h$ is the Planck constant and $v_{\rm DM}$ is the DM particle velocity, being itself associated to the fluctuation size for matter waves.  $\lambda$ becomes smaller than the Earth diameter for $m_{\rm DM} \gtrsim 10^{-10}$ eV, for galactic velocities.

There are multiple theoretical frameworks, like ultra-light axions or fuzzy DM, in which DM form compact mini-halos, clumps or solitonic cores, with a mass between $10^9$ kg and $10^{15}$ kg.  For instance, in fuzzy dark matter, solitonic cores can form with a comparable size to the Compton wavelength, which can be much smaller than the Earth's size.  Those types of models have recently attracted more attention due to the absence of WIMP detection in colliders and in direct or indirect detection experiments.
Dark matter clustering or inhomogeneities on very small scales is therefore an interesting possibility. Setting new limits on this clustering would help to distinguish various plausible dark matter candidates.  
Another possibility that is relevant for our work is the formation of compact dark matter objects such as boson clouds or axion stars, which could also have a mass in the asteroid-range depending on the model.   Black holes formed in the very early Universe, called primordial black holes (PBHs), is another possibility discussed in more details hereafter.

\subsection{PBH and existing limits}

PBHs may have formed through various mechanisms including the collapse of order one density fluctuations produced during inflation or reheating, phase transitions or topological defects, see e.g.~\cite{LISACosmologyWorkingGroup:2023njw,Carr:2023tpt,Carr:2020xqk} for reviews of recent developments.  The detection of gravitational-waves and the intriguing properties of compact binary coalescences have rekindled the idea that a significant fraction or even the totality of the DM could be made of PBHs~\cite{Bird:2016dcv,Clesse:2016vqa,Sasaki:2016jop}. Contrary to black holes of stellar origin, there is no intrinsic limitation on the PBH mass other than from their evaporation through Hawking radiation and from the current size of the Hubble horizon.   But there exist a lot of astrophysical and cosmological limits on their abundance.  In the context of this work, we are interested in light, asteroid-mass PBHs.  These are interesting because there is still an open window between $10^{14}$ kg and $10^{20}$ kg where the abundance of PBHs still remains unconstrained\footnote{See however~\cite{Esser:2022owk} for model-dependent limits or~\cite{Smirnov:2022zip} for observational hints in this range, coming from the destruction of stars in dwarf galaxies by captured PBHs}.  At lower mass, between $10^{11}$ kg and $10^{16}$ kg, all the limits repose on the black hole evaporation process, for instance based on gamma-ray sources or on the energy injection in the primordial plasma that should have significantly impacted the temperature anisotropies of the cosmic microwave background.   However, as recently pointed out in~\cite{Auffinger:2022dic}, all these limits still have important model dependencies and uncertainties.  Taking them into account, it is not excluded that PBHs significantly contribute to the DM for masses above approximately $10^{13}$ kg.  This is without considering that the black hole evaporation itself is still a theoretical conjecture that has not been proven observationally.  The Hawking radiation can also depend on the underlying fundamental theory.    

As a consequence, it is highly interesting to search for complementary probes of asteroid-mass PBHs that are not relying on the Hawking evaporation mechanism.  This is the case here because the signal from near-Eearth PBHs in GNSS and gravimeters is solely induced by the gravitational Newtonian force. 

\subsection{Dark matter in the solar system}\label{subsec:DM_in_the_SS}

Previous works dealing with the gravitational detection of DM objects in the solar system relied on the simple approximation a constant object flux.  Despite its simplicity, this approach can nevertheless provide a first and useful estimate as well as an upper bound for the expected number of detections for a given sensitivity of GNSS and gravimeters.  One can then derive plausible bounds on the DM clump, compact object or PBH abundance for each technique, in case of absence of observation.  

The number of DM objects $N_{\rm o}$ of mass $m_{\rm o}$ passing at a distance $d$ of the center of the Earth during a survey duration $T_{\rm s}$ can be calculated using the mass flux relation:
\begin{equation}\label{eq:N_objects_basics}
    N_{\rm o} \approx n_{\rm o} \, \pi d^2 \, V_{\textrm{DM}}\, T_{\rm s} \, ,
\end{equation}
where $V_\textrm{DM}$ is the relative velocity of the DM object w.r.t to a specific reference frame, in our case the Earth-centered inertial (ECI) coordinate frame.  $n_o$ is the number of density of dark objects,
\begin{equation}\label{eq:n_objects_basics}
    n_{\rm o} = \frac{ f_{\rm o} \rho_{\textrm{DM}, \odot}}{m_\textrm{\rm o}},
\end{equation}
where $f_{\rm o}$ is the DM density fraction made of such dark objects and $\rho_{\textrm{DM},\odot}$ is the DM density in the Solar System neighbourhood, estimated from the rotation curve of the Milky Way to be about \cite{Salucci:2010qr,Bovy:2012tw}:
\begin{equation}\label{eq:rho_DM}
\rho_{\textrm{DM},\odot}  \simeq  0.4\, \textrm{GeV/cm}^3 \simeq 0.009\, M_{\Sun} {\rm pc}^{-3}.
\end{equation}
The mass range of interest for our analysis will be $m_\textrm{DM} = [10^{10}, \ 10^{20}]$ kg, for which one could expect more than one event for a survey duration of several years and at a distance typically larger than the size of the Earth radius but still smaller than the typical interplanetary distance, as discussed below. 


The DM velocity in our galactic halo can be described by a Maxwellian-Boltzmann distribution that is relatively independent of the galactocentric radius.  
The associated probability density function is therefore given by
\begin{equation}\label{eq:Vdm_distribution}
    f_{V}(V_\textrm{DM}) = \sqrt{\frac{2}{\pi}}\, \frac{1}{\Delta^3}V_\textrm{DM}^2e^{-V_\textrm{DM}^2/(2\Delta^2)},
\end{equation}
where $\Delta$ is the velocity dispersion, related to the rms-velocity for DM of about $V_{\rm rms} \approx 220$ km/s by $\Delta = V_{\rm RMS}/\sqrt{3}$ \cite{Xu2008}. 

\subsection{g-force variations}
\label{subsec:Basics_of_signal}
\begin{figure}[t]
    \centering
    \includegraphics[width=0.47\textwidth]{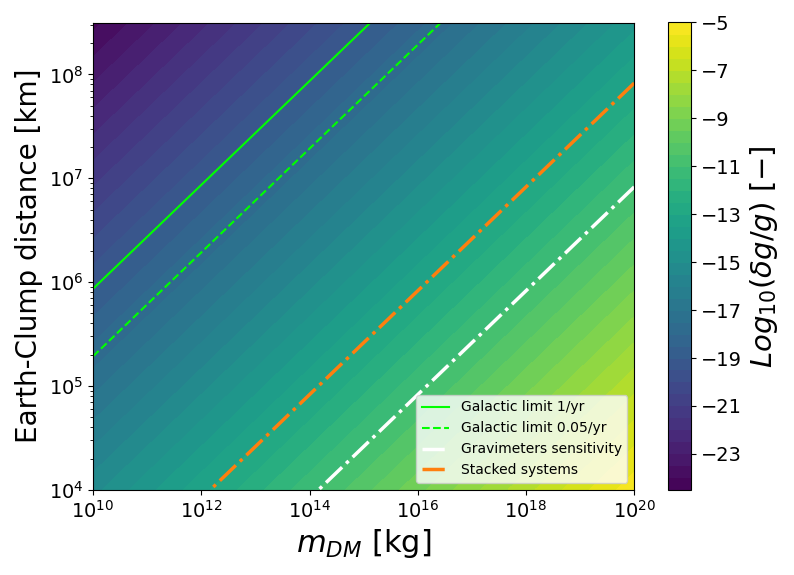}
    \caption{Estimation of the change in local gravity intensity $\delta g/g$ as a function of the DM clump mass, $m_\textrm{DM}$, and the Earth-Clump distance. The approximate ``single-probe'' sensitivities of the gravimeter ($10^{-11}$) systems are displayed, in comparison to the ``galactic limit'' based on the simple estimate of the flux assuming one event per year (solid green line) and one per 20 years (dashed green line). The sensitivity offered by GNSS systems (not displayed) is lower in this quick estimate, which will be contradicted by the in-depth study in Section 5. The stacked system (orange dashed line) represents the possible sensitivity after a statistical analysis based on the cross-correlation over the GNSS constellations and a dozen gravimeters.}
    \label{fig:dg_plot_plot}
\end{figure}

The typical normalised change in local gravitational acceleration, $\delta g/g$, due to a near-Earth DM object, can be estimated by assuming that its gravitational force on the sensor, here a GNSS satellite or a ground-based gravimeter, is aligned with the Earth gravitational force:
\begin{equation}\label{eq:dg_g_basics}
   \left\vert \frac{\delta g}{g} \right\vert = \left\vert \frac{\delta F_{\rm G}}{F_{\rm G}} \right\vert = \frac{m_\textrm{DM}}{M_\Earth} \ \left(\frac{r_{\Earth/\rm{s}}}{R}\right)^2 \, ,
\end{equation}
where $F_G$ is the gravitational force, $M_{\Earth}$ is the mass of the Earth, $r_{\Earth/\rm{s}}$ is the  Earth-sensor distance and $R$ is the distance of the DM object to the Earth center. This approximate expression is valid as long as $R \gg r_{\Earth/\rm{s}}$.
\autoref{fig:dg_plot_plot} shows $\delta g/g$ as a function of the object mass and nearest distance, assuming a velocity $V_\textrm{DM} = 200$ km/s, based on the simple estimation of Eq.~(\ref{eq:dg_g_basics}).  The isocontours for which $N_{\rm o} = 1$ assuming $T_{\rm s} = 1$ yr and 20 yr are also displayed and can be compared to roughly estimated GNSS and gravimeter sensitivities introduced in Section \ref{sec:Introduction}.
%


The dashed orange line in Fig.\ref{fig:dg_plot_plot} represents the possible sensitivity obtained after a full statistical analysis based on cross-correlation between a hundred GNSS satellites and a dozen superconducting gravimeters around the world. Using these simple estimations, the current sensitivities are about four orders of magnitude too low for hoping a direct post-detection in the available data. 
However, this simple estimate is also highly conservative and neglects important effects that are investigated in our paper.  First, it is based on Eq.~(\ref{eq:rho_DM}) for the local DM density in the object mass flux.  However, this simple estimate does not consider the full distribution of velocities and the possible enhancement of the DM density close to the Earth due to (1) the gravitational focusing by the Sun or (2) the possible capture through the Solar System bodies. Hence, to determine the event rate, we will compute in Section \ref{sec:Trajectories} the fly-by distance $d$ and $V_\textrm{DM}$ using a semi-analytical approach for the DM clump trajectories, and will obtain the phase space density close to the Earth.
We will then model the shape of the GNSS and gravimeter signals to obtain a better determination of the sensitivity level. 
%
%
\subsection{Constraints by spacecrafts and planetary ephemeris}\label{sec:ephemeris}

Constraints on the distribution of DM in the solar system exist in the case of WIMP clouds, which would have a continuous influence on planetary ephemerides and spacecraft positions \cite{Iorio2010,Pitjev:2013sfa}, where uncertainty is dominated by the lack of knowledge about the mass of the asteroid belt. Let's start from the naive assumption of directly transposing constraints on the density of 'WIMPS' obtained from planetary ephemerides to constraints on the density of dark matter clumps of asteroidal mass. 
Then, the permanent presence of a cloud of WIMPs should be equivalent to a very episodic presence of asteroid-mass clumps in hyperbolic (or highly elliptical) orbits, so a transient phenomena by nature. In this case, the constraints provided by planetary ephemerides would be in fig.\ref{fig:dg_plot_plot} stronger than for an individual probe (gravimeter or satellite), but weaker than for a statistical correlation on a ensemble of gravimeters or a GNSS constellation over several years of data.

Actually, we can assert that the current constraints on WIMPS dark matter in the solar system are not directly applicable to transient phenomena for several reasons.
First the applicability of the constraints on DM in the solar system obtained by planetary ephemeris is still a matter of debate. Indeed the usual method consists in the simple calculation of the additional perturbative acceleration induced by DM mass. However, a recent study \cite{Fienga:2023ocw} pointed that the proper procedure would be to use the planetary ephemeris to constrain a hypothetical DM distribution in the solar system directly at the level of the numerical integration of the equations of motion. This would imply to fit the data to the modeled ephemeris using a statistical analysis, and interpret the significance of the residual. Indeed, the various correlations between the parameters have to be intrinsically taken into account.

The main principles of such an analysis have been recently described in \cite{Tran:2023jci}. They simulated the long-term variation of the distance between the Earth and other planets in the inner solar system due to a transitory passage of a PBH within a few astronomical units. They estimate a potential detection sensitivity of one event per decade, for PBHs with a mass greater than $10^{16}$ kg, making their study complementary to ours.

In addition, measurements of the density of dark matter clouds are based on 'snapshots' of the situation in the Solar System, on the scale of 100 years. However, the transit frequency of dark matter clumps of asteroidal mass remains very low, with very small masses involved. Indeed, we show in this paper that the constraints we project to obtain after a full correlation analysis implies a state of equilibrium with a total DM clump mass of 0.001 Ceres mass in a 1.5 AU (Astronomical Unit) sphere around the Sun, for individual DM clumps endowed with a mass between $10^{8}$ and $10^{17}$ kg.
A much larger time scale is therefore needed, of the order of the solar system's lifetime, for the multiplication of transits to be considered as equivalent to a permanent cloud of clumps. In this way, constraints could be imposed by models of the evolution of the solar system over the long term, including dark matter clump transits. 
To our knowledge, there's no guarantee that the constraints given by such a study would be meaningful, as they would be plagued by major uncertainties about the solar system evolution model.

\section{Hyperbolic trajectories of unbound objects}
\label{sec:Trajectories}
In order to create a large Monte-Carlo synthetic population, a semi-analytical approach of a DM object fly-by is used. The orbit characteristics are defined by the following Keplerian elements: the semi-major axis $a$, the eccentricity $e$ and the true anomaly $\theta$. In addition, for the three-dimensional orientation of the orbit, one needs the Right-Ascension of the Ascending Node (RAAN) $\Omega_\textrm{DM}$, the argument of the periapsis $\omega_\textrm{DM}$ and the inclination $i_\textrm{DM}$. 

\subsection{Semi-analytical model}
%

Only two parameters are required to fully determine the two-dimensional hyperbolic orbit:
\begin{itemize}
    \item The impact parameter: $B$
    \item The hyperbolic excess velocity: $V_{\infty}$.
\end{itemize}
The impact parameter is the minimum distance between the main celestial body and the asymptote of the hyperbolic trajectory. The excess velocity is reached by the body as the distance from the Sun tends to infinity.

The semi-major axis $a$ for the hyperbolic trajectory is the distance from the crossing point of asymptotes to the periapsis. This parameter is defined as negative:
\begin{equation}\label{eq:semimajor_axis_hyp}
    a = -\frac{\mu_{\Sun}}{V_{\infty}^2} \, ,
\end{equation}
where $\mu_{\Sun} = G\, M_{\Sun}$ is the Sun's standard gravitational parameter. The eccentricity is then calculated as:
\begin{equation}\label{eq:ecc_hyp}
    e = \sqrt{1 + 2  \frac{h^2 \mathcal{E}}{\mu_{\Sun}^2}} \, ,
\end{equation}
where $h$ is the specific orbital angular momentum which is computed as function of $B$ and $V_\infty$:
\begin{equation}\label{eq:angular_momentum_spec}
    h \approx B \ V_{\infty} \, .
\end{equation}
In eq. (\ref{eq:ecc_hyp}), $\mathcal{E}$ denotes the specific mechanical energy which is also a function of orbital parameters,
\begin{equation}\label{eq:energy_spec}
    \mathcal{E} =  \frac{1}{2}V_{\infty}^2 - \frac{\mu_{\Sun}}{r_{\infty}} \,   ,
\end{equation}
where $r_{\infty}$ is the initial condition simulating the infinite distance to the Sun.
%
%
We use the concept of Sphere of Influence (SOI) \cite{Wakker2015} to estimate $r_{\infty}$:
\begin{equation}\label{eq:r_SOI}
    \bar{r}_{{\rm SOI,}\Sun} \approx 0.923\, \left\lvert a_m \right\rvert \, \left( \frac{m}{M}\right)^{2/5} \, ,
\end{equation}
where $a_m$ is the semi-major axis of the smaller mass $m$ orbiting the larger mass $M$. The infinite radius is then: $r_{\infty} \approx \bar{r}_{{\rm SOI,}\Sun}$. The combination of the Sun's mass $M_{\Sun}$, the mass of the closest galactic mass and the distance from the sun to the galactic center \cite{Xu2008} enables to determine the solar system SOI, resulting in $r_{\infty} \sim 2 \cdot 10^{15}$ m. Hence, the excess velocity is restricted, $V_{\infty}>0.16$ km/s, following (\ref{eq:energy_spec}).

Given the three Keplerian elements ($a$, $e$ and $\theta$) defined within the orbital plane, the radial distance to the Sun is then finally computed as:\begin{equation}\label{eq:keplerian_radius}
    r(\theta) = \frac{a(1-e^2)}{1+e \ \text{cos}(\theta)} \, .
\end{equation}
The following expression provides the true anomaly as a function of time:
\begin{equation}\label{eq:theta_dot_eq}
    \theta(t)-\theta_0 = \int_0^t \frac{\sqrt{\mu_{\Sun} \ a(1-e^2)}}{r^2(t')}\ dt' \, ,
\end{equation}
where $\theta_0$ is the initial true anomaly.
Finally, additional transformations are required from the perifocal coordinate system to the Sun's equatorial frame to orient the orbit in the Cartesian three-dimensional space. 

\subsection{Closest approach}\label{sec:flux_cloest_point}
\begin{figure}
    \centering
    \includegraphics[width=0.48\textwidth]{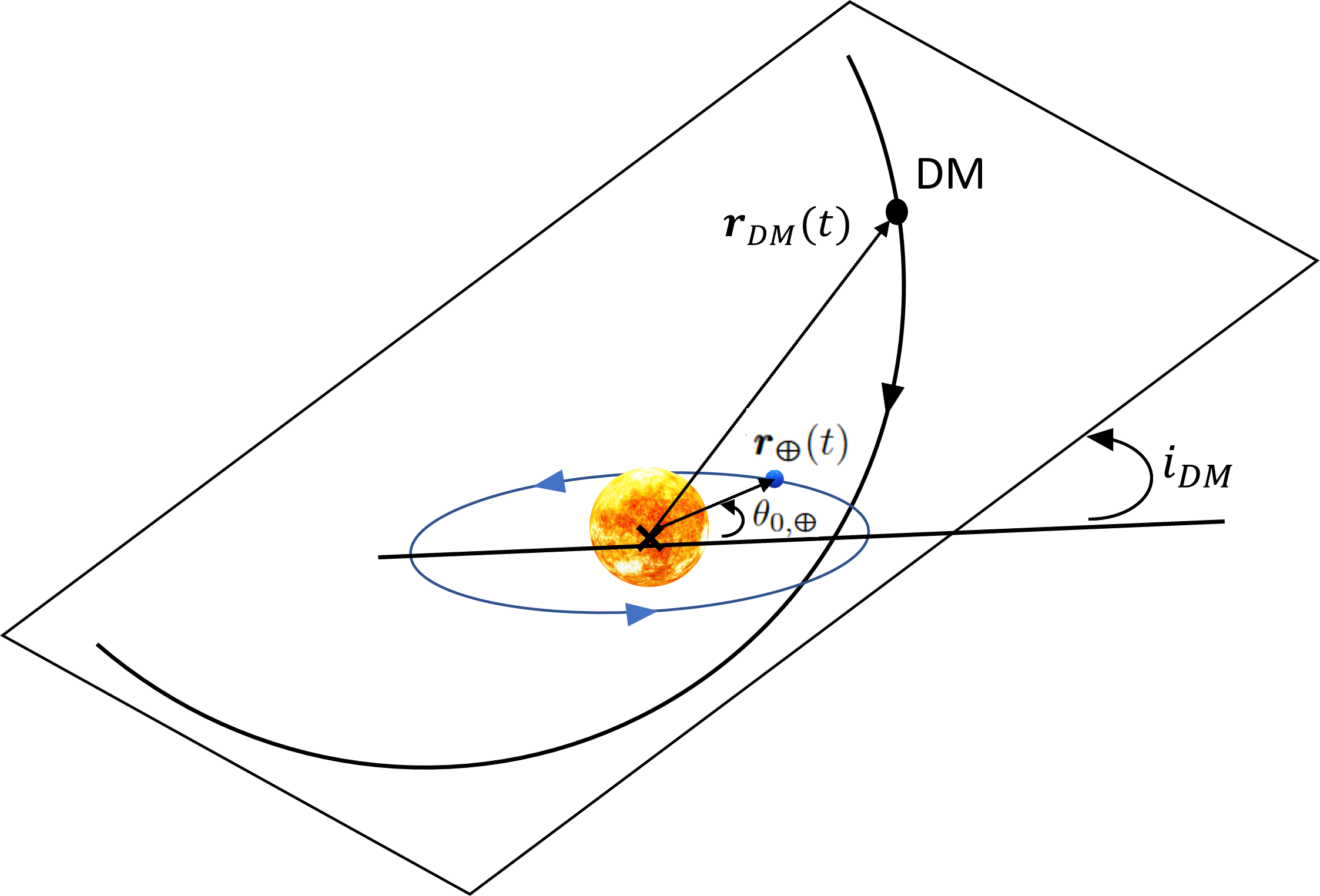}
    \caption{Three dimensional simplified representation of DM heliocentric fly-by $\pmb{r}(t)$ with the inclination angle $i_\textrm{DM}$. The Earth $\Earth$ orbit is also modeled with the line of apsides (black line) defining $\theta_{0, \Earth}$.}
    \label{fig:simplified_3d_orbits}
\end{figure}

The closest distance approach $d$ of the DM clump w.r.t Earth is computed as the minimum of $|| \pmb{r}(t) - \pmb{r}_{\Earth}(t)||$, with $\pmb{r}_{\Earth}(t)$ being the Earth's position vector w.r.t the Sun (see \autoref{fig:simplified_3d_orbits}). For the DM clump, the position and velocity vectors are first computed in PQW-coordinates (perifocal coordinates) centered on the Sun as:
\begin{equation}
    \pmb{r}(t) = r(t) \, [\text{cos}(\theta)\, \hat{\pmb{p}} + \text{sin}(\theta)\, \hat{\pmb{q}}],
\end{equation}
\begin{equation}\label{eq:V_DM_sun}
    \pmb{V}(t) = \frac{\mu_{\Sun}}{h}\cdot [-\text{sin}(\theta(t))\hat{\pmb{p}} +  (e + \text{cos}(\theta(t)))\hat{\pmb{q}}],
\end{equation}
where the unit vector $\hat{\pmb{p}}$ is directed towards the periapsis of the orbit. Then, the position and velocity vectors are translated into a Sun-Centered Inertial (SCI) Cartesian coordinate system using:
\begin{equation}\label{eq:transformation}
    \text{R}_{PQW \rightarrow SCI} = \text{R}_z(\Omega_\textrm{DM})\, \text{R}_x(i_\textrm{DM})\,  \text{R}_{z}(\omega_\textrm{DM}) \, ,
\end{equation}
where $\text{R}_z$ and $\text{R}_x$ are the unit rotation matrices. A similar transformation has been done for the Earth position and velocity vectors.

Finally, the relative velocity $V_{\textrm{DM}/ \Earth}$ is simply computed at the specific time of closest approach as:
\begin{equation}\label{def:VDM/E}
V_{\textrm{DM}/ \Earth} = || \pmb{V}_{\textrm{DM}/ \Sun} - \pmb{V}_{\Earth/ \Sun}|| \, .     
\end{equation}

\subsection{Input and output parameters}
\label{subsec:inputs}

\begin{figure}[h]
    \centering
    \includegraphics[width=0.50\textwidth]{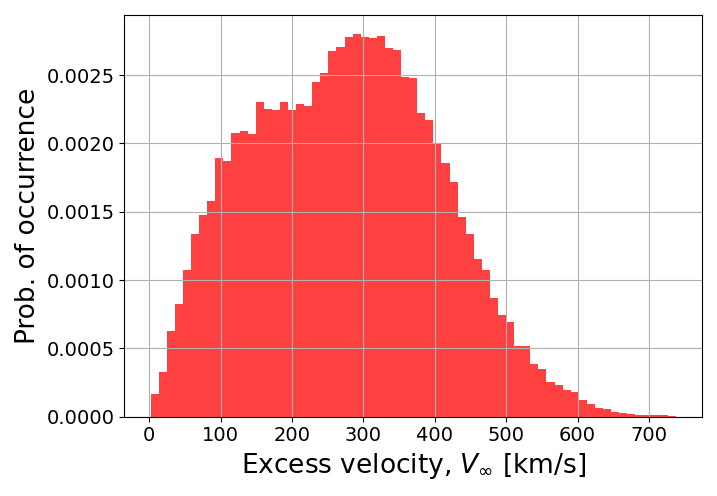}
    \caption{Pseudo-Maxwellian Distribution for $V_{\infty}$ ($N_{\rm tot} = 4\cdot 10^4)$.}
    \label{fig:vinf_dist}
\end{figure}

To estimate the closest approach of the DM clump w.r.t Earth, a Keplerian hyperbolic orbit is constructed for every ($B$, $V_{\infty}$) pair and ensemble of Keplerian angles $i_\textrm{DM},\ \omega_\textrm{DM}$ and $\Omega_\textrm{DM}$. As there is no particular reason for the DM clumps to have certain values for those angles, they are all uniformly distributed as: $i_\textrm{DM}\sim \mathcal{U}[0, \pi]$ rad and  $\Omega_\textrm{DM},\omega_\textrm{DM}\sim \mathcal{U} [0, 2\pi]$ rad. 
%
%
Using (\ref{eq:Vdm_distribution}), the distribution of hyperbolic excess velocities (relative speed of DM w.r.t the Sun) is obtained from: 
\begin{displaymath}
V_{\infty} \approx \sqrt{V_\textrm{DM}^2 + V_{\Sun}^2 - 2V_\textrm{DM} V_{\Sun}\cdot \text{cos}(\sigma)},
\end{displaymath}
where $V_{\Sun} \sim 208$ km/s \cite{Xu2008} is the Sun's velocity w.r.t the galactic center and $\sigma$ is the relative angle between $\pmb{V}_\textrm{DM}$ and $\pmb{V}_{\Sun}$ (uniformly distributed: $\sigma\sim \mathcal{U}[0,\ \pi]$). The DM velocity, $V_{\rm DM}$ is distributed with a Maxwellian-Boltzmann distribution as explained in \autoref{subsec:DM_in_the_SS}. The resulting new distribution for $V_{\infty}$ is denoted in this study as ``Pseudo-Maxwellian" visualised in \autoref{fig:vinf_dist}.  

%

These Keplerian initial condition of the DM clumps can be translated to a Cartesian state vector $[\pmb{r}_{\infty}, \pmb{V}_{\infty}]^T$. Additionally, a circular orbit for the Earth is simulated from an initial condition $\theta_{0, {\Earth}}$ taken as: $\theta_{0, \Earth} \sim \mathcal{U}[0,2\pi]$. The reference for this angle is taken from the line of apsides, represented as the black line in \autoref{fig:simplified_3d_orbits}. The set of used variables for the Monte-Carlo simulations and desired output are summarised in \autoref{tab:variables_1} of Appendix \ref{app:Monte-Carlo}, with additional technical details.

Let us assume that the obtained minimum distance $d$ is distributed around the input value $B = 1$ AU, without taking into account the gravitational focusing by the Sun. Thus, in order to focus on the events of interest, namely the smallest observed distances with $d < 1$ AU (referring to fig. \ref{fig:dg_plot_plot}), an additional simulation is performed with $B \sim \mathcal{U}[0.98, 1.025]$ AU. The probability $P_B$ for this case scenario to happen is
\begin{displaymath}
P_B(B_\textrm{max},B_\textrm{min}) = \frac{0.045}{B_\textrm{max}-B_\textrm{min}}\, P_B(0.98, 1.025)\, , 
\end{displaymath}
compared to the more general case with $B_{max} = 100$ AU and $B_{min} = 0.01$ AU. A similar approach is followed for the inclination distribution for which, $i_\textrm{DM} \sim \mathcal{U}[0^\circ,0.05^\circ]$:  
\begin{displaymath}
P_i(i_\textrm{max},i_\textrm{min}) = 2 \frac{0.05^\circ}{i_\textrm{max}-i_\textrm{min}}\, P_i(0^\circ, 180^\circ)\, , 
\end{displaymath}
where the factor 2 in the right hand side reflects that a retrograde orbit $i_\textrm{DM} \sim 180^\circ$ provides the same result as $i_\textrm{DM} \sim 0^\circ$.

\subsection{Event-rate Estimations}\label{subsec:event_rates_results}

\begin{figure}[h]
    \centering
    \includegraphics[width=0.48\textwidth]{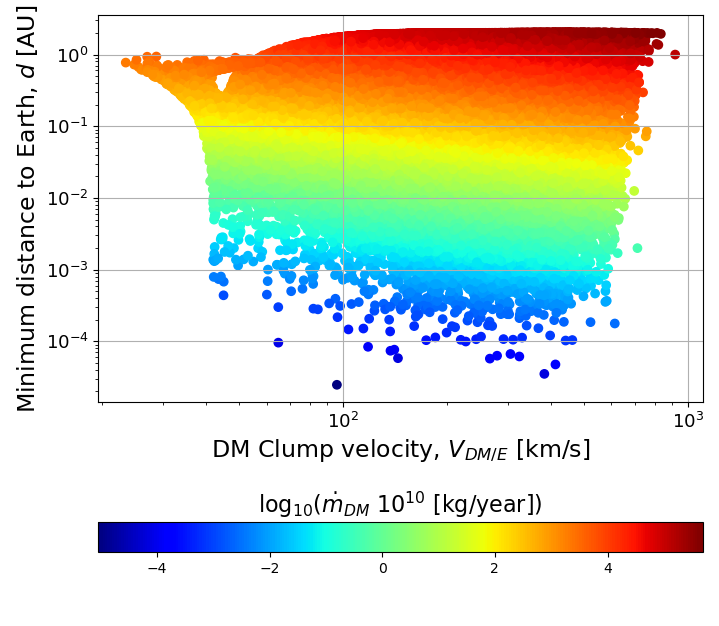}
    \vspace{-0.30cm}
    \caption{Resulting distribution from Monte-Carlo simulation of minimum Earth-Clump distance $d$ and velocity $V_{\text{DM}/ \Earth}$, with mass flow $\dot{m}_\textrm{DM}$ heat-map, for the constraining assumption on the impact parameter ($B \sim \mathcal{U}[0.98, 1.025]$ AU).}
    \label{fig:m_dot_year}
\end{figure}

\autoref{fig:m_dot_year} shows the output of the MC simulation in terms of the minimum distance $d$ to the Earth and the velocity distribution $V_{\mathrm{DM}/ \Earth}$, see (\ref{def:VDM/E}). The associated mass flux $\dot{m}_\textrm{DM}$ is presented as a heat-map. Given that a model of point mass has been introduced, the notion of mass flux is independent on the actual mass distribution of the DM clump in the Standard Galactic halo. For example, \autoref{fig:m_dot_year} shows that a mass flux of $10^{10}$ kg/year has a close encounter with the Earth at a distance of about $2\, 10^{-3}$ AU and a relative velocity around 500 km/s. This means that these events frequencies with these given fly-by characteristics are equivalent:
\begin{itemize}
    \item a (single) clump with a mass of $10^{10}$ kg every year,
    \item a clump with a mass of $10^{12}$ kg every century,
    \item a clump with a mass of $8.3\, 10^8$ kg every month,
\end{itemize}
and cannot be distinguished without an underlying model for DM halo fragmentation, see section \ref{sec:theory_and_simple_estimates}.

A gap of points around a velocity of $45$ km/s and a distance of $0.5$-$0.7$ AU is observed in \autoref{fig:m_dot_year}, which interestingly is close to the solar system escape velocity at the Earth's orbital distance, namely 42.4 km/s.

\begin{figure}[h]
    \centering
    \includegraphics[width=0.40\textwidth]{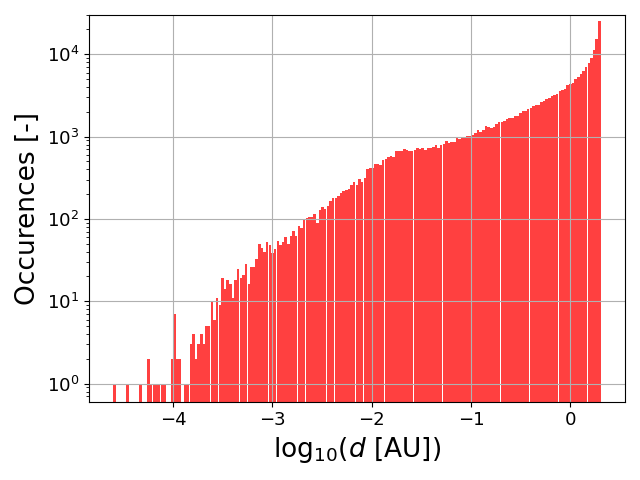}
    \caption{Histogram of the minimum distance $d$ distribution from Monte-Carlo simulation for the constraining assumption on the impact parameter ($B \sim \mathcal{U}[0.98, 1.025]$ AU).}
    \label{fig:d_distribution}
\end{figure}

\autoref{fig:d_distribution} shows the histogram for the minimum distance $d$ distribution for the constraining assumption on the Sun impact parameter $B \sim \mathcal{U}[0.98, 1.025]$ AU. 
Some clump trajectories intercept the orbital sphere of GNSS constellations, with value of $d$ less than $2\cdot 10^{-4}$ AU, for a probability of occurrence $N_\textrm{event}/N_\textrm{total}$ of about $10^{-4}$. These close approaches are associated to velocities within the range of $60-500$ km/s as can been seen in \autoref{fig:m_dot_year}. 
This is due to the fact that these velocities exceed the $30$ km/s orbital speed of the Earth, allowing the clump to overtake it more easily. 
%
%
Combining \autoref{fig:m_dot_year} and \autoref{fig:d_distribution}, fly-by events intercepting the orbital sphere of GNSS constellations represent a mass flux of $10^7$ kg per year. 

Finally, the influence of the inclination angle $i_{\rm DM}$ on the distribution of $d$ is described in Appendix \ref{app:Monte-Carlo} and turns out to be negligible.



\subsection{Sun gravitational focusing}

\begin{figure}[h]
    \centering
    \includegraphics[width=0.42\textwidth]{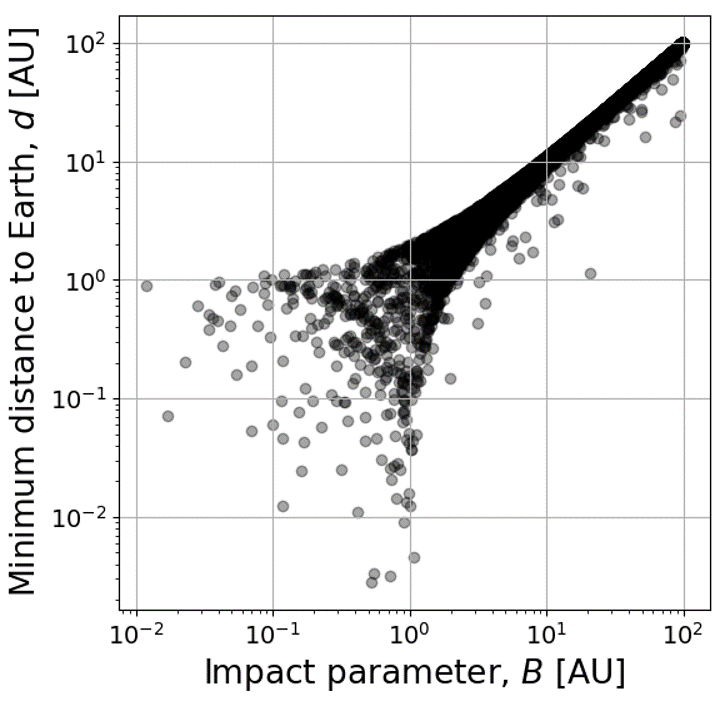}
    \caption{Minimum Earth to Impact parameter distance as a function of the heliocentric impact parameter with the broader distributions.}
    \label{fig:d_over_B_func_B}
\end{figure}

Knowing the orbital properties of DM cumps or PBHs, it is possible to estimate the enhancement in the number of near-Earth trajectories due to the gravitational focusing by the Sun.  From the MC simulations with a wide uniform prior for $B$ and with the full distribution for DM velocities $V_{\rm DM}$, in particular from the distribution of the minimum distance to Earth $d$ as a function of $B$ shown in \autoref{fig:d_over_B_func_B}, one expects that trajectories significantly impacted by the gravitational focusing so that they lead to a close approach are very rare.  But the ones with $B$ very close to 1 AU leading to a close approach are also rare, so it is not straightforward to infer the importance of each case with respect to the other and it is worth to further investigate the relevance of the gravitational focusing effect.  

\begin{figure}[h]
    \centering
    \includegraphics[width=0.40\textwidth]{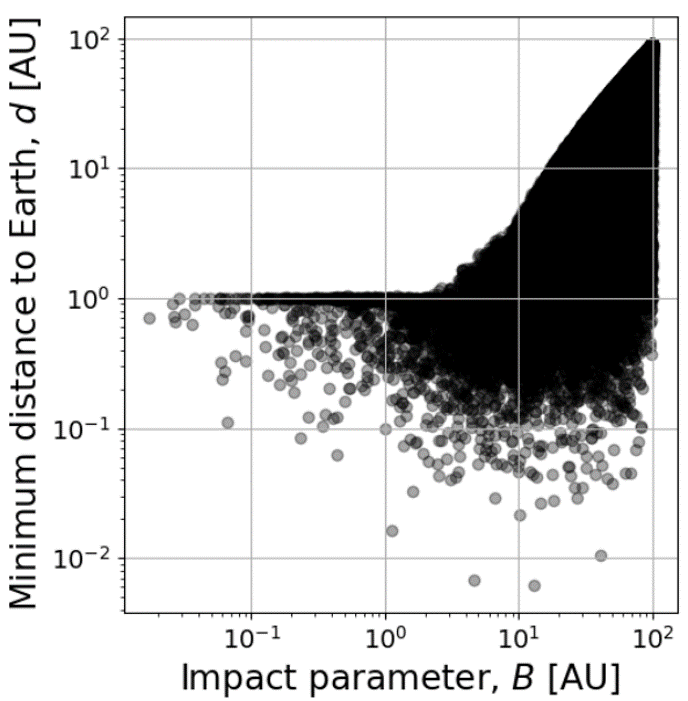}
    \caption{Minimum Earth to Impact parameter distance as a function of the heliocentric impact parameter with the small $V_{\infty}$-distribution.}
    \label{fig:d_func_B_small_Vinf_interval}
\end{figure}

For this purpose, we have performed another MC simulation focusing on small velocities, with a restricted uniform prior on $V_{\rm DM}$ between 0.5 and 10 km/s, for which the gravitational focusing is typically important.  The results are shown in \autoref{fig:d_func_B_small_Vinf_interval}.  They show that only a small fraction of trajectories lead to $d<0.1$ AU with impact parameters between 1 and 100 AU and only two trajectories over $10^6$ lead to $d<0.01$ AU.  Given that the MC is restricted to about a $\mathcal O(10^{-2})$ fraction of all possible trajectories, this gives an extremely low probability, of order $10^{-6}$, that the gravitational focusing leads to $d< 0.01$ AU.  For comparison, we obtain a probability of order $10^{-6}$ in the case where the gravitational focusing effect is irrelevant.  

\begin{figure}[h]
    \centering
    \includegraphics[width=0.5\textwidth]{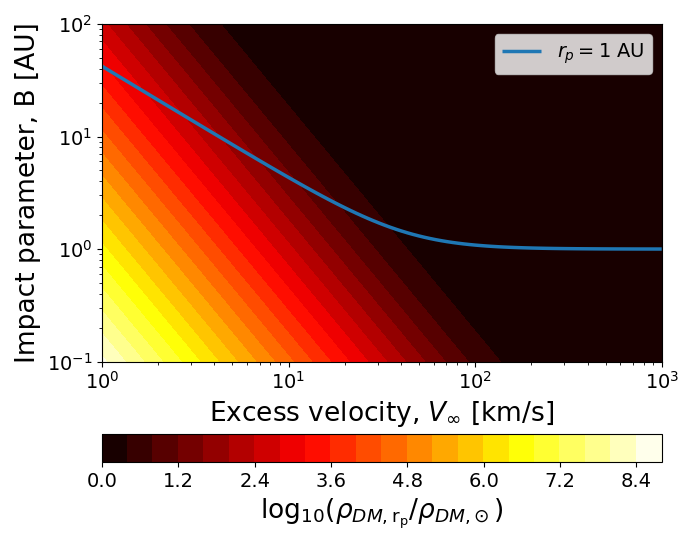}
    \caption{Local density enhancement factor as a function of the impact parameter and excess velocity with the pericenter as the chosen characteristic distance.}
    \label{fig:rho_DM_factor}
\end{figure}

Alternatively, for a small interval around a given $B$, one can calculate, for all the possible values of $V_{\infty}$, the enhancement in the DM  density at the trajectory perihelion 
\begin{equation}\label{eq:pericenter}
    r_{\rm p} = \frac{\mu}{V_{\infty}^2}\left[ \sqrt{1 + \frac{B^2 V_{\infty}^4}{\mu^2}}-1\right],
\end{equation}
due to the gravitational focusing effect.  This enhancement goes like $\rho_{\rm DM, r_p}/ \rho_{\rm DM, \Sun} = (B / r_{\rm p})^2 $, simply due to the reduction of the flux section area ($r_{\rm p} \propto B$ when the gravitational focusing effect is significant).  This enhancement factor is shown in \autoref{fig:rho_DM_factor}.  In this figure, we also represent the line corresponding to $r_{\rm p} = 1$ AU, i.e. on the Earth orbit.  A significant enhancement factor between 10 and 40 can be obtained, but only for very low values of $V_\infty \lesssim 5 {\rm km/s}$ that would correspond to less than a percent of all possible trajectories (see \autoref{fig:vinf_dist}). Therefore, the concerned number of trajectories is so low that it largely compensates the enhancement in density. 

In summary, using two arguments based on MC simulations and on the estimation of the possible DM density enhancement due to the gravitational focusing, one can conclude that this effect remains small and cannot boost the expected number of near-Earth trajectories for galactic DM objects.   


\section{Dark Matter capture in the solar system}\label{sec:DM_capture}

Different capture mechanisms of interstellar DM objects might have strongly enhanced the DM density in the solar system.  A capture can result from the energy loss of a  DM object due to the momentum transfer in a 3-body problem involving the Sun, one of the planets and the DM object itself.  The deflection of a trajectory can be calculated by assuming that the DM object mass is negligible compared to the planet mass.  
As a preamble, it's worth mentioning that any capture requires $V_{\infty} \lesssim 40$ km/s \cite{Napier2021} considering Jupiter as the third body. 
This already shows that only objects with Galactic velocities in the lower end of the Maxwellian distribution can be captured.
In what follows, we will consider separately the deviation of a trajectory by the Sun or by one of the planets of the solar system.


\subsection{Deflection by the Sun}

A direct capture by the Sun is possible due to the Sun motion around the Solar System barycenter. The deflected orbital velocity $V_{\rm f}$ of a DM object caused by the Sun having a velocity $U_{\Sun}$ w.r.t the Solar System barycenter is given by~\cite{Napier2021}
\begin{equation*}
    V_{\rm f}^2  = V^2 + 4 \, U_{\Sun}^2 - 4 \, U_{\Sun} \, V \text{cos}(\beta_\infty)
\end{equation*}
where $\beta_\infty$ is the deflection angle and $V$ is the ``unperturbed'' heliocentric velocity in a hyperbolic orbit with no energy transfer effect caused by the Sun's movement around the Solar System barycentre, given by
\begin{equation}
    V(r)^2 = V_{\infty}^2 + \frac{2\, \mu_{\Sun}}{r}.
\end{equation}
The deflection angle is related to the eccentricity of the two-body problem initial hyperbolic trajectory through $\cos(\beta_\infty) =1/e$ and $U_{\Sun}$ can be related to a fraction of the orbital velocity $V_{\rm J}$ of Jupiter, being with a mass $M_{\rm J}$ the main contributor to the Sun motion, 
with an additional numerical factor $\alpha \approx 0.21$ \cite{Napier2021},
\begin{equation}
    U_{\Sun} = 2\alpha  \frac{M_{\rm J}}{M_{\Sun}}\, V_{\rm J}\, .
\end{equation}
A capture is then be characterised by the condition $V_{\rm f} < V_{\rm esc}$, where $V_{\rm esc} = \sqrt{2\mu_{\Sun}/r}$ is the solar system escape velocity.  To evaluate if the DM clump has been captured by the Sun, a condition on the difference between final deflected velocity and escape velocity, $\Delta V_{\rm f}(r) = V_{\rm f}(r)-V_{\rm esc}(r) <0$ is checked over all possible orbital radii $r$.
When a DM object is directly captured by the Sun, the bounded orbit has a semi-major axis $a_{\rm b}$ given by
\begin{equation}\label{eq:a_b}
    a_{\rm b} =- \left( \frac{V_{\rm f}(r)^2}{\mu_{\Sun}} - \frac{2}{r}\right)^{-1}.
\end{equation}

\subsection{Three-body capture due to a planet}

The second capture mechanism consists in a strong deflection of the trajectory by a planet, leading to an elliptical orbit for the DM clump of which one of the focal point is the Solar System barycenter. Our calculation of the velocity deflection due to a planet (P) flyby is mainly based on \cite{Xu2008} and consists on producing a synthetic Monte-Carlo population of analytical DM clump hyperbolic trajectories. Therefore, no numerical simulations are implemented within this section of the study. Consequently, due to the analytical nature of the method, additional probabilistic corrections are required and will be described below. Furthermore,  this study adds to the methodology in \cite{Xu2008} by correctly computing the real 3D nature of the DM and planetary orbits (with the angles $i_\textrm{DM}$, $\Omega_\textrm{DM}$ and $\omega_\textrm{DM}$). Two other methods \cite{Khriplovich2009, Napier2021} can also be found in the literature. In the study of Khriplovich et al. \cite{Khriplovich2009}, the methodology is completely performed analytically, in which the mass estimation is derived from analytical and probabilistic laws. Lastly, Napier et al. \cite{Napier2021, Napier2021_2} perform both orbital numerical simulations and analytical derivations of the capture cross-section of interstellar objects. Thus, by simplifying the numerical nature of orbital simulations, this study provides a faster approach in estimating DM density in the Solar System. 

First, the DM clump velocity in a frame at rest with the Sun can be found using \autoref{eq:V_DM_sun}, in which
$\theta$ is the true anomaly at the planet-Sun distance, denoted by $a_{\rm P}$, $\hat{\pmb{p}}$ and $\hat{\pmb{q}}$ are the perifocal coordinates (PQW) unit vectors within the DM clump's orbital plane. The planet's orbit is then assumed to be circular, leading to the following planet velocity vector:
\begin{equation}\label{eq:V_P_sun}
    \pmb{V}_{{\rm P}/\Sun} = V_{\rm P} \, [-\text{sin}(\theta_{\rm P})\, \hat{\pmb{p}}' +  \text{cos}(\theta_{\rm P})\, \hat{\pmb{q}}']\, ,
\end{equation}
where $\theta_{\rm P}$ is the true anomaly of the planet, taken as a randomly distributed angle between 0 and $2\pi$, $\hat{\pmb{p}}'$ and $\hat{\pmb{q}}'$ are the perifocal coordinates unit vectors within the planet's orbital frame. 
We obtain the expression of $\hat{\pmb{p}}'$ and $\hat{\pmb{q}}'$ in the DM perifocal coordinate system $\hat{\pmb{p}}'$ and $\hat{\pmb{q}}'$ using the \autoref{eq:transformation} and its inverse transformation with the DM orbit inclination angles.

To facilitate the computation of the final heliocentric velocity $\pmb{V}_{\rm f}$, the DM's radial, along-track and cross-track reference frame (RSW) is used leading to\cite{Xu2008}:
\begin{equation}
\pmb{V}_{\rm f} =
\begin{bmatrix} 
    V_{{\rm DM}_{\parallel}}\text{cos}(\beta_{\rm P}) + V_{{\rm DM}_{\perp}}\text{sin}(\beta_{\rm P}) \\
    -V_{{\rm DM}_{\parallel}}\text{sin}(\beta_{\rm P}) + V_{{\rm DM}_{\perp}}\text{cos}(\beta_{\rm P}) \\
    0
\end{bmatrix}
- \pmb{V}_{{\rm P}/\Sun}|_{RSW}
\end{equation}
where $V_{DM_{\parallel}}$ and $V_{DM_{\perp}}$ are the DM clump parallel and perpendicular to the radial direction velocities in the planet's rest-frame and $\beta_{\rm P}$ is the deflection angle caused by the planet. These two velocities components and deflection angle can be computed using the procedure presented in \cite{Xu2008}. Lastly, the capture condition similar to the one for the Sun is checked:  \begin{equation} \label{def:planet_capture}
    ||\pmb{V}_{\rm f}|| < V_{\rm esc}(r = a_{\rm P}) = \sqrt{\frac{2\mu_{\Sun}}{a_{\rm P}}}.
\end{equation} 
The resulting bounding orbit has a semi-major axis $a_{\rm b}$ which can be computed using \autoref{eq:a_b} in the case of $r = a_{\rm P}$. 

The Monte-Carlo simulation of an ensemble of orbits provides only the probability that an orbits fulfills \autoref{def:planet_capture} at the sphere of radius $a_{\rm P}$ centered on the Sun. 
This occurrence probability of captured events is denoted as $P_{\rm c, MC}$.
In addition, it is necessary to add corrections (not performed in \cite{Xu2008}) to the result obtained, which take into account simplified analytical probability laws. 
%
%

The first probability relates to the crossing of the DM clumps' orbit with the planet's orbit around the Sun. Under the assumption of a uniform distribution of impact parameters, this fraction denoted as $f_{{\rm P}_1}$, can be approximated as:
\begin{equation}
    f_{{\rm P}_1} \approx \left( \frac{B_{\rm max}}{2 r_{\infty}} \right)^2\, ,
\end{equation} where $B_{\rm max}$ is computed as in \cite{Napier2021}:
\begin{equation}
    B^2_{\rm max} = a_{\rm P}^2+ 2|a| a_{\rm P} \, .
\end{equation}

Furthermore, in addition to the fractional probability of crossing the planet's orbit, only a small fraction of DM objects actually enters the planet's SOI. This fraction $f_{{\rm P}_2}$ can be  approximated as \cite{Napier2021}:
\begin{displaymath}
    f_{{\rm P}_2} \approx \left( \frac{r_{\rm SOI, P}}{2 a_{\rm P}} \right)^2 \, .
\end{displaymath} 


Combining all the above, the corrected probability of captured objects, $P_{\rm c}$, can be computed as:
\begin{equation}\label{def:P_c}
    P_{\rm c} = P_{\rm c, MC} \ f_{{\rm P}_1} \ f_{{\rm P}_2}\, .
\end{equation}


\subsection{Capture and ejection rates}\label{subsec:capture_rates_results}

Assuming that the Solar System is moving in the DM halo throughout its existence, the total flux of DM mass crossing the solar system since its creation reads:
\begin{equation}\label{eq:M_DM}
    M_\textrm{DM} \sim  \pi \bar{r}_{\infty}^2\, \rho_{\textrm{DM},\odot}\, V_{\Sun}\, \Delta t_{\rm SS} \cos(\theta_{\rm SS}) \left( 1+ \frac{\bar{V}_{\rm esc}^2}{\bar{V}_{\infty}^2} \right) \, ,
\end{equation}
where $\Delta t_{\rm SS}\approx 1.4\cdot 10^{17}$ s is the Solar System's age and $\theta_{\rm SS} \approx 30^\circ$ is the complementary angle to the angle between the ecliptic plane and the galactic plane. The last term in \autoref{eq:M_DM} approximates the Sun's gravitational focusing with the average escape velocity $\bar{V}_{\rm esc} \sim 165.3$ km/s, taken over a radial distance interval [$10^{-3}, 10^{4}$] AU, and the average excess velocity $\bar{V}_{\infty} = 274$ km/s (see \autoref{fig:vinf_dist}). Given all the aforementioned estimates, $M_\textrm{DM} \sim 134\ M_{\Sun}$ (compared to $203M_{\Sun}$ in \cite{Xu2008}).

Each planet contributes to the capture by a fraction:
$M_{\rm c} \approx M_\textrm{DM}\cdot P_{\rm c}$. \autoref{tab:masses_results} shows the capture probabilities $P_{\rm c}$ and mass estimates $M_{\rm c}$ computed in this study in comparison with results obtained previously in the Literature.
\begin{table*}[ht]
\centering
\caption{Capture probability $P_{\rm c}$ per planet with total mass captured $M_{\rm c}$ in the Solar System, in comparison with the values found in the Literature \cite{Xu2008, Khriplovich2009}}
\label{tab:masses_results}
\begin{tabular}{m{2.0cm}m{2.5cm}m{2.7cm}m{3.5cm}}
\hline
\textbf{Planet} & $P_{\rm c}$  [-]     & $M_{\rm c}$ [kg]  &  \textbf{\cite{Xu2008},\cite{Khriplovich2009} values} [kg] \\ \hline
Earth & $3.0\cdot 10^{-18}$ &  $5.9\cdot 10^{16}$    & $3.8\cdot 10^{17}$, $3.9\cdot 10^{15}$\\ 
Jupiter & $1.8\cdot 10^{-14}$ & $3.5\cdot 10^{18}$  & $4.9\cdot 10^{19}$, $1.2\cdot 10^{18}$\\ 
Saturn & $1.3\cdot 10^{-14}$ & $2.5\cdot 10^{18}$   & $2.8\cdot 10^{19}$, $3.7\cdot 10^{17}$\\ 
Uranus  & $6.4\cdot 10^{-15}$ & $1.2\cdot 10^{18}$ & $1.3\cdot 10^{19}$, $5.7\cdot 10^{16}$\\ 
Sun  & $< 10^{-15}$ & $< 1.8\cdot 10^{17}$ & $-$\\ 
\hline
\end{tabular}
\end{table*} 
With regards to \cite{Xu2008, Khriplovich2009}, the results found in this study are one order of magnitude more conservative relative to \cite{Xu2008}, but less conservative in comparison with \cite{Khriplovich2009}. These discrepancies are described in detail in Appendix \ref{app:Capture}.  
The order of magnitude of the total mass captured by the planets since the solar system formation is about $8\cdot 10^{18}$ kg, in comparison to the value of $2\cdot 10^{18}$ kg obtained in \cite{Khriplovich2009}.
%
%
The captured mass by the Sun is constrained by $M_{\rm c} < 10^{17}$ kg, making its contribution negligible.

The input data of our semi-analytical simulation for the computation of closed orbits and their properties resulting from the capture process by Solar System bodies are described in the Appendix \ref{app:Capture}. The results of the simulation show that DM clumps or PBHs follow a strongly elliptical orbit around the Sun after their gravitational capture, due to their high velocities. Indeed, the eccentricity of the simulated orbits $e_{\rm b}$ exceeds 0.91 whereas the majority of captured DM objects have pericenter values below 1 AU, with a mean of $\approx 0.3$ AU, resulting in possible Earth close approaches. However, the captured object have orbital periods far beyond the human lifetime due to the high semi-major axis.

%



In addition, the ejection rate is a significant concern as found in \cite{Edsjo2010,Malyshkin1998,Peter2009} and would then lead to a high decrease of the DM population in the Solar System. Given that the pericenter $r_{\rm p}$ of the new bound orbit is well below the orbital distance of Jupiter, the chance of ejection of the DM clump is considerably increased \cite{Napier2021_2}.
%
The survival fraction $f_{\rm s}(t)$ of \cite{Napier2021_2} can be used for that purpose and leads to $f_{\rm s}(\Delta t_{\rm SS}) \approx 10^{-5}$, given the age of the solar system. The resulting equilibrium or steady-state mass is in the range: $1.149-4.594\cdot 10^{13}$ kg for our semi-analytical method for the capture rate. This result is close to the steady-state estimate by \cite{Napier2021_2} of $10^{14}$ kg, using purely numerical methods. Therefore, our assessment of the additional density in the Earth's vicinity reads:
\begin{equation}
\Delta \rho_\textrm{DM} \sim 10^{13}/(4\pi/3 (1\text{AU})^3) \approx 2\rho_\textrm{DM}    
\end{equation}
which does not represent a significant increase in density.

\section{Expected Signatures}\label{sec:expected_signatures}

We propose to use gravimeters and GNSS satellites as two complementary networks of sensors to probe passing DM clumps. For gravimeters, a perturbation in the gravitational field could be measured, coming from the radial component of the Newtonian force induced by a third-body, whereas for satellites, the observable is a perturbation in the nominal reference orbit.

The signature of a DM clump flyby is again obtained with semi-numerical simulations of the clump hyperbolic orbit, based on \autoref{eq:keplerian_radius}. However, the calculation is now performed in the Earth-centered frame.  We denote by $\pmb{r}_\textrm{DM}(t)$ the position of the clump relative to Earth when \autoref{eq:theta_dot_eq} is used to compute the clump trajectory. The parameters $(V_{\infty}, B, \theta_0)$ are again used to define the hyperbolic trajectory in the orbital plane, but now they denote the impact parameter, the excess velocity and the initial true anomaly relative to Earth.  Finally $(i_\textrm{DM}, \ \omega_\textrm{DM}, \ \Omega_\textrm{DM})$ define the orbital plane w.r.t the Earth's celestial sphere in Cartesian coordinates.

\subsection{GNSS}
The Global Navigation Satellite Systems (GNSS) consist in constellations of around 30 satellites at an altitude of about 20 thousands kilometers with a revolution period of a bit more than ten hours.  The four global constellations are Galileo (EU), GPS (USA), GLONASS (Russia) and BeiDou (China).  GNSS rely on precision timing signals provided by on-board atomic clocks. The clock synchronization errors and the precise orbits are routinely computed with a high degree of accuracy and are publicly released by the International GNSS Service (IGS)~\cite{Johnston2017}.

\subsubsection{Signal modelling for one satellite}

The expected signature of a DM clump on a GNSS satellite orbit is obtained by computing the Newtonian acceleration of the satellite induced by such a third-body, and by determining the resulting three-dimensional Keplerian osculating orbit as a function of time, based on \autoref{eq:keplerian_radius}.  Initially, the GNSS orbit is determined by $(a,\ e, \ i, \ \omega, \ \Omega, \ M_0)_{\rm GNSS}$, where $M_0$ is the initial mean anomaly.  The true anomaly is then calculated with \autoref{eq:theta_dot_eq}, whereas \autoref{eq:ecc_hyp} is used as a check.

Given these orbital paths, the orbit perturbation on the GNSS is then computed as a third-body perturbation acceleration leading to:
\begin{equation}\label{eq:da_dg_GNSS}
    \frac{\delta \tilde{\pmb{a}}(t)}{g(t)} = - \frac{\mu_{\textrm{DM}}}{g(t)} \left(\frac{\pmb{r}-\pmb{r}_{\textrm{DM}}}{|| \pmb{r}-\pmb{r}_{\textrm{DM}}||^3} + \frac{\pmb{r}_{\textrm{DM}}}{||\pmb{r}_{\textrm{DM}}||^3}\right)\, ,
\end{equation}
where $\pmb{r}$ is the position (w.r.t Earth) of the GNSS satellite and $g(t)$ is the reference gravity field value for the GNSS constellation. The last term in (\ref{eq:da_dg_GNSS}) comes from the fact that the non-inertial reference frame centered on the Earth is in free fall in the clump gravitational potential. This global term is identical for all GNSS satellites.

\begin{figure*}[ht]
    \centering
    \includegraphics[width =1.0\textwidth]{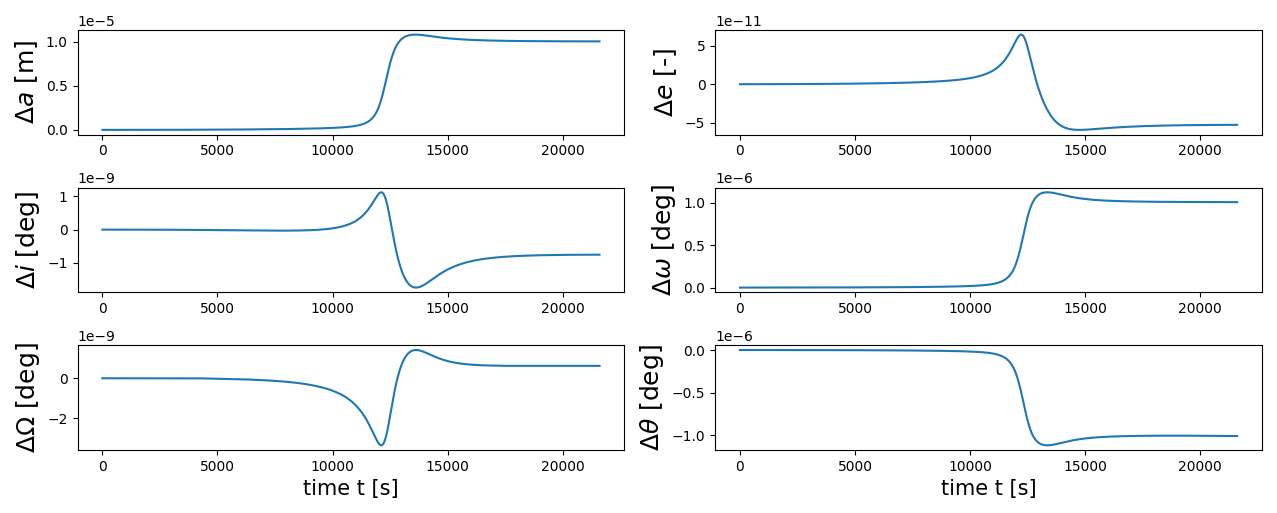}
    \caption{Single satellite reaction to gravitational impulse for a clump ($10^{15}$ kg) passage with $B=10000$ km, $V_{\infty} = 15$ km/s and $i_\textrm{DM} = 0$ deg (with a half-day simulation time and $\Delta t = 1$ s).}
    \label{fig:observables_GPS}
\end{figure*}
The deviation of the GNSS reference orbit due to the aforementioned perturbation is performed by solving the Keplerian elements using Gauss' variational equations \cite{Aslanov2017, DeRuiter2013}:
\begin{equation}\label{eq:a_dot}
    \frac{\textrm{d}a}{\textrm{d}t} = \frac{2\, a^2}{h^2} \left[ e\, \text{sin}(\theta)\, \tilde{a}_r + \frac{p}{r}_0 \, \tilde{a}_{\theta}\right]  ,
\end{equation}
\begin{equation}
    \frac{\textrm{d}e}{\textrm{d}t} = \frac{1}{h} \left\{\tilde{a}_r\, p\, \text{sin}(\theta) + \tilde{a}_{\theta} \left[e\, r_0 + (p+r_0)\cos(\theta)\right] \right\},
\end{equation}
\begin{equation}
    \frac{\textrm{d}\theta}{\textrm{d}t} = \frac{h}{r_0^2} - \frac{1}{e\, h} \left[ -\tilde{a}_r\, p\, \cos(\theta) + \tilde{a}_{\theta}(p+ r_0)\sin(\theta)\right],
\end{equation}
\begin{equation}
    \frac{\textrm{d}i}{\textrm{d}t} = \tilde{a}_z \, \frac{r_0}{h} \cos(\theta + \omega),
\end{equation}
\begin{equation}
    \frac{\textrm{d}\omega}{\textrm{d}t} = -\frac{\textrm{d}\theta}{\textrm{d}t} +\frac{h}{r_0^2} - \tilde{a}_z \frac{r_0}{h}\cot(i) \, \sin(\theta +\omega),
\end{equation}
\begin{equation}
    \frac{\textrm{d}\Omega}{\textrm{d}t} = \tilde{a}_z \, \frac{r_0}{h} \frac{\sin(\theta + \omega)}{\sin(i)}\, ,
\end{equation}
where $\pmb{r}_0 (t)$ is the initial reference orbit, $p$ is the orbital semilatus rectum: $p=h^2/\mu$, $\mu$ here refers to the Earth's celestial parameter and $\tilde{a}_r$, $\tilde{a}_{\theta}$ and $\tilde{a}_z$ are the radial, tangential and vertical components (within RSW satellite frame) of the perturbation acceleration. These represent the response of the satellite to a transient input and provide actual observables.
Since a singularity is encountered in the present case of a (quasi) circular orbit, the relations are transformed according to those found in \cite{Khalil} with modified Keplerian elements $e_x$ and $e_y$, so that $e_y/e_x = \tan(\omega))$ and $u = \theta + \omega$.
%
Considering the typical orbital parameters of a Galileo satellite, an outline of the reaction of one satellite is shown in \autoref{fig:observables_GPS}. The overall nature of the output is a step, with only the change in true anomaly being unstable and periodical, potentially due to the insufficient orbital stiffness of the satellite.

\subsubsection{Signal modelling for the Galileo satellites constellation}
\label{subsec:gnss_signal_res}
\begin{figure}[h]
    \centering
    \includegraphics[width =0.5\textwidth]{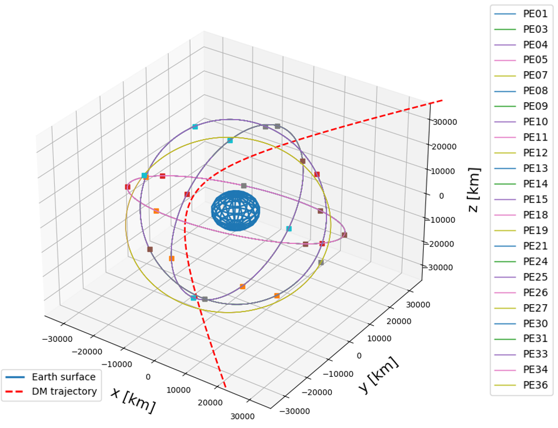}
    \caption{Numerically simulated DM and Galileo constellation orbits with the final position of the satellites.}
    \label{fig:orbits_GALILEO}
\end{figure}

GNSS constellations will simply be modeled using the same methodology as above, and tracking the various orbital positions. The data to simulate the constellation orbits are taken from \cite{Galileo_orbits} with the Galileo constellation as case study. Currently, the Galileo constellation includes 24 active satellites placed in MEO orbits with semi-major axis of 29599.8 km. As illustrated by our simulation in \autoref{fig:orbits_GALILEO}, the Galileo satellites are distributed into 3 orbital planes with an nominal inclination of 56°, distributed evenly round the equator according to three different Right-Ascensions of the Ascending Node, RAAN. These satellites on a same orbital plane are equally spaced by various phase differences in the true anomaly.

\begin{figure*}[ht]
    \centering
    \includegraphics[width =0.75\textwidth]{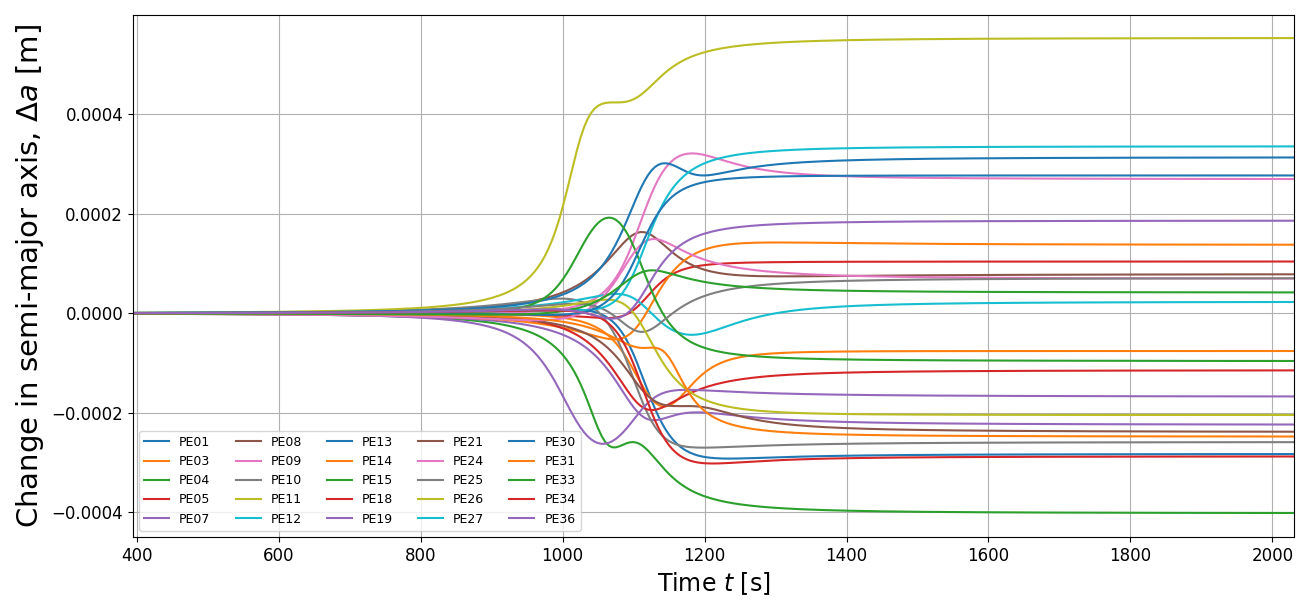}
    \vspace{-0.3cm}
    \caption{Impulse response to the semi-major axis of the entire Galileo constellation for a clump ($10^{15}$ kg) passage with $B=15500$ km, $d=14495.5$ km, $V_{\infty} = 300$ km/s and $i_\textrm{DM} = 90$ deg.}
    \label{fig:delta_a_c_ex}
\end{figure*}
The signature of a DM clump flyby in \autoref{fig:delta_a_c_ex} is obtained by combining the modeling of the DM clump orbit, the simulation of the orbits of the various satellites and the gravitational perturbation induced by the DM clump (\ref{eq:da_dg_GNSS}).
This forms the basis for observing the correlation between the signals from the different probes when considering different scenarios related to the incidence direction, the closest approach distance and the excess velocity of the DM clump.
In \autoref{fig:delta_a_c_ex}, the semi-major axis has been chosen as the reference Keplerian element for data analysis, as it offers a direct estimate of distance and energy, see section \ref{sec:Trajectories} translating the Earth as the main attractor: $\mu_\Earth$. It is also the orbital parameter whose variation can be measured with the best precision, see section \ref{sec:Sensitivity}.

Let us briefly analyse the signal of \autoref{fig:delta_a_c_ex} which simulates the flyby of a DM clump $10^{15}$ kg with an (Earth) impact parameter $B=15500$ km, $V_{\infty} = 300$ km/s and $i_\textrm{DM} = 90$ deg. The impulse responses form an envelope, with  half of the constellation having a negative settling, due to the geometric orbital distribution of the satellites in \autoref{fig:orbits_GALILEO}. The simulation also shows a slight delay between different satellites and a larger impact on the orbit whenever the DM clump passes close to the satellite. Finally, it is essential to mention that the entire constellation acts nearly as one system. If an event occurs, it would be detected for all satellites. Therefore, the limits of the envelope provide the associated fly-by masses that could be observed. 

\subsubsection{Sensitivity analysis for different parameters}
In this subsection, we visualise how the semi-major axis $a$ is sensitive to a given change in the DM orbital parameters, focusing on the effect of $B$ and $V_{\infty}$. This is important for future statistical data analysis, e.g. to known if the satellites react in unison, or if there is a delay conducive to time correlation analysis.

\begin{figure}[h]
    \centering
         \includegraphics[width=0.475\textwidth]{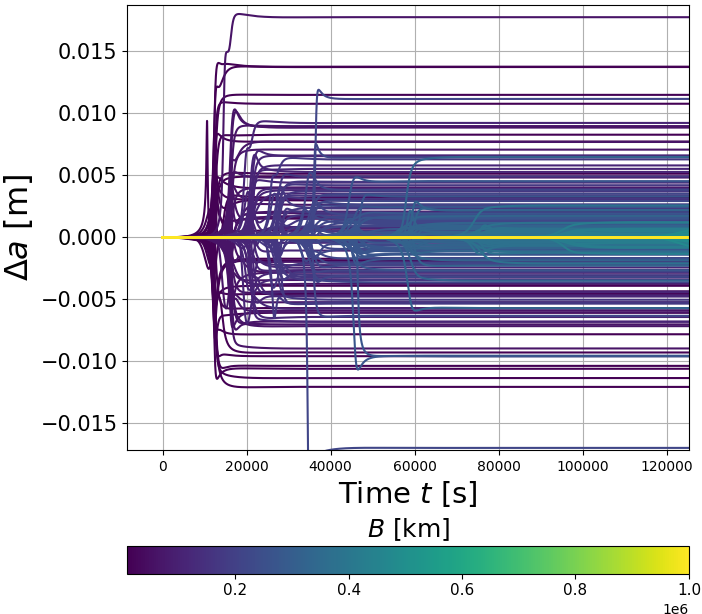}
        \caption{Entire Galileo constellation signal variation as a function of $B$ with $V_{\infty}$ and $i_\textrm{DM}$ fixed at 15 km/s and 90$^\circ$.}
        \label{fig:gnss_sens_B}
\end{figure}

The sensitivity in terms of $B$ is driven by the fact that $da/dt$ is proportional to the acceleration in \autoref{eq:a_dot}, with a caveat for $B<30000$ km. In that case, the DM clump can be observed twice by the constellation. Indeed, two effects are combined in equation \ref{eq:da_dg_GNSS}: 
\begin{enumerate}
    \item The earth center recoil, common to all Galileo satellites and maximum at the moment of Earth closest approach.
    \item The delayed satellite closest approach, different for each Galileo satellite.
\end{enumerate}
Hence the complex double-step pattern seen in \autoref{fig:delta_a_c_ex} or in \autoref{fig:gnss_sens_B} for lower values of $B$. 
For $B>30000$ km, the system comprising the Earth and Galileo satellites can be considered as a point object from the DM clump point of view. Hence the complex time pattern gradually disappears for larger value of $B$ in \autoref{fig:gnss_sens_B}. In parallel, the effect is weaker but the event duration is longer, losing the characteristic step profile.

\begin{figure}[h]
         \centering
        \includegraphics[width=0.45\textwidth]{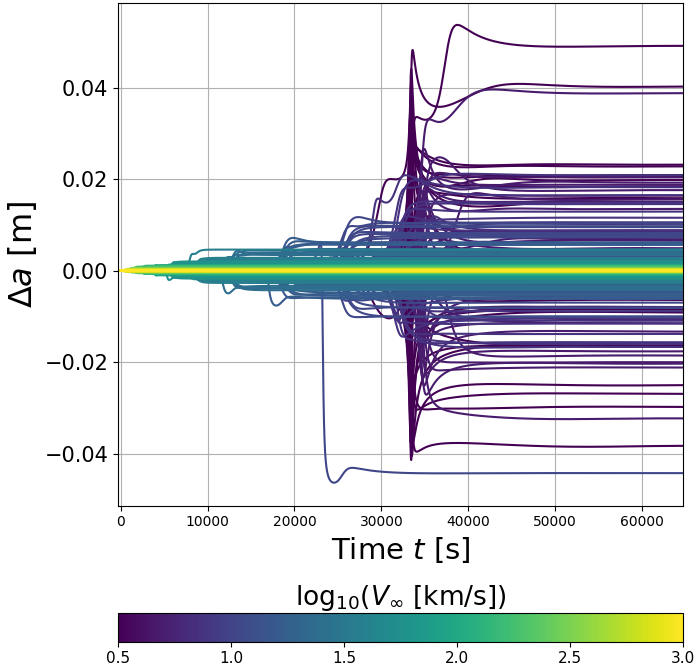}
        \caption{Entire Galileo constellation signal variation as a function of $V_{\infty}$ with $B$ and $i_\textrm{DM}$ fixed at 15500 km and 90 deg.}
        \label{fig:gnss_sens_V_inf}
\end{figure}

For what concern the effects of $V_{\infty}$ shown in \autoref{fig:gnss_sens_V_inf}, the key element when $B < 30000$ km is the duration of the constellation crossing. Indeed, the double-step pattern appears narrower as speed increases while the amplitude of the effect decreases. In the case where $B < 30000$ km, the change in the acceleration profile is nearly the same for the entire constellation and becomes independent of $V_{\infty}$. 
%
%


\subsection{Gravimeters}

In this section, we will model the signature of the DM clump fly-by over a network of 9 stations provided with superconducting gravimeters, distributed around the world. In the future, this would allow, through time-correlations and template-based searches, to seek for the synchronous occurrence and the propagation of glitches in the gravitational potential.

\subsubsection{Earth crossing}

The Earth crossing occurs when the Earth closest approach of the hyperbolic orbit is smaller than the Earth's radius or: $d < R_{\Earth}$. For the sake of simplicity, it will be assumed in this Section that DM interacts with ordinary matter only via the gravitational force, and not via the electromagnetic nor strong or weak nuclear forces.

A method similar to that of \cite{Horowitz2019} will be used to estimate the gravitational signal in the case of a DM cluster passing inside the Earth. The gravimeters should then feel two distinct effects. We ignore in this analysis the tidal effect of the DM clump generating on the Earth's interior structure. The first one is due to the Earth's center of mass recoil, $d^2X_{\Earth}/dt^2$, as Earth falls in the DM clump gravitational potential. The second effect is due to the actual acceleration component on the gravimeter caused by the DM clump, denoted as $g_\textrm{DM}$.

The first effect can be approximated as follows:
\begin{equation}\label{eq:d2X_dt2}
    \frac{\textrm{d}^2 X_{\Earth}}{\textrm{d}t^2} \approx \frac{m_\textrm{DM}}{M_{\rm enc}(r_\textrm{DM})}\cdot \left( \frac{\textrm{d}^2r_\textrm{DM}}{\textrm{d}t^2} - r_\textrm{DM}\, \dot{\theta}^2\right) \, ,
\end{equation}
where it is assumed that $m_\textrm{DM}<<M_{\rm enc}$ with $M_{\rm enc}(r)$ being the Earth's enclosed mass at the radial distance $r_\textrm{DM}(t)$, and $\dot{\theta}(t)$ is the time derivative of the true anomaly. 

The enclosed mass is computed using the Preliminary Reference Earth Model (PREM) \cite{Dziewonski1981, PREM_data}, which assumes that the Earth is spherical and that the density, $\rho_{\Earth}$, only depends on the radial distance from the center. This enclosed mass is then: 
\begin{equation}\label{eq:M_enc}
    M_{\rm enc}(r) = \int_0^{r} 4\pi r^2 \rho_{\Earth}(r')\ {\rm d}r'.
\end{equation}
\autoref{eq:M_enc} shows that the capture by the Earth of a DM object which crosses its surface seems highly unlikely since a strong gravitational pull would be essential to slow down highly energetic DM clumps. Additionally, any resulting elliptical orbit that intersects the Earth's surface would require highly fine-tuned initial conditions to be stable over longer periods of time. This, therefore, shows the issues with the premise presented in \cite{Horowitz2019}, in which a DM clump would have been captured by Earth and would have a circular orbit inside of it.

\begin{figure}[h]
    \centering
    \includegraphics[width =0.38\textwidth]{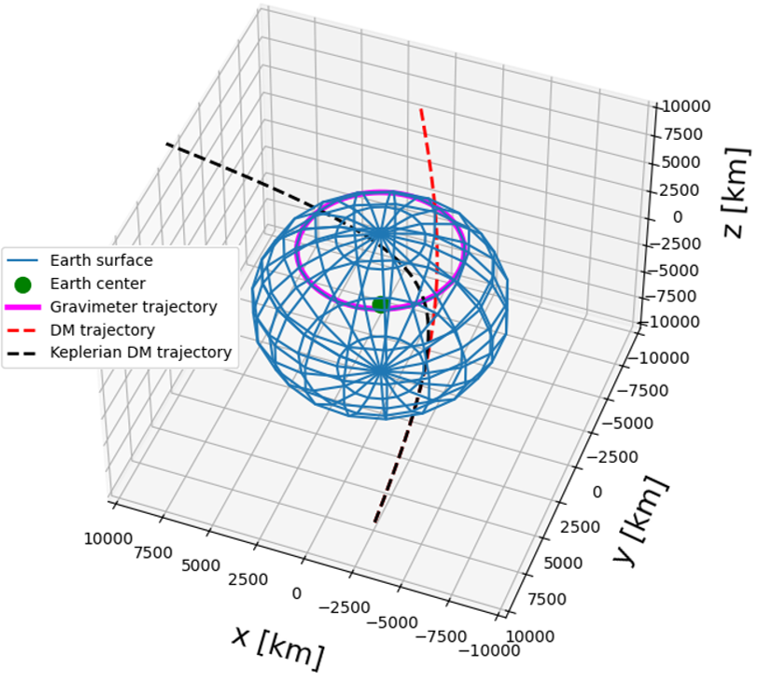}
    \caption{Numerically simulated DM and gravimeter trajectories.}
    \label{fig:orbits_gravi1}
\end{figure}

Although the hyperbolic orbit is assumed Keplerian, the DM clumps will be gradually less deflected as it enters inside of the Earth, since $M_{\rm enc}(r) \leq M_{\Earth}$ for all $r\leq R_{\Earth}$. A Runge-Kutta 4 (RK4) numerical integration was used to simulate the orbit from the moment it enters the Earth with:
\begin{equation}\label{eq:r_DM_Earth-crossing}
    \frac{{\rm d}^2 \pmb{r}_\textrm{DM}}{{\rm d}t^2} = - G\, \left[M_{\rm enc}(r_\textrm{DM})+m_\textrm{DM}\right] \, \frac{\pmb{r}_\textrm{DM}}{r_\textrm{DM}^3}.
\end{equation}
The associated trajectory can be visualised and compared to the assumption of a point mass for the Earth in \autoref{fig:orbits_gravi1}. The DM clump initially follows the classical Keplerian orbit, then becomes less and less deviated inside the Earth in comparison with the point-mass model, as predicted by Gauss's law.

Assuming that the DM clump is a point mass relative to the size of the Earth, the radial acceleration $g_{{\rm DM}, r}$ induced by the DM clumps reads as follows:
\begin{equation}
    g_{{\rm DM},r} = -\mu_\textrm{DM}\, \frac{\pmb{r}_{\rm g} - \pmb{r}_\textrm{DM}}{||\pmb{r}_{\rm g} - \pmb{r}_\textrm{DM}||^3} \cdot \frac{\pmb{r}_{\rm g}}{||\pmb{r}_{\rm g}||}
\end{equation}
where $\pmb{r}_{\rm g}$ is the position of the gravimeter. This position is computed as follows \cite{Namigata2022,Horowitz2019}: 
\begin{equation}\label{def:r_g}
\pmb{r}_{\rm g} = R_{\Earth} \cos(\theta_\ell) [\text{cos}(\omega_{\Earth}t+\Lambda_\ell), \text{sin}(\omega_{\Earth}t+\Lambda_\ell),\tan(\theta_\ell)]^T \, ,   
\end{equation}
where $\theta_\ell$ and $\Lambda_\ell$ are the latitude and longitude of the gravimeter and $\omega_{\Earth}$ is the Earth's rotational rate, as illustrated in the simulation of \autoref{fig:orbits_gravi1} (purple line).

Finally, the time-dependent normalised gravimeter reading, $\delta g_r(t)/g$ is calculated with:
\begin{equation}
    \frac{\delta g_r(t)}{g} = \frac{1}{g} \,\left[ g_{{\rm DM}, r} + \frac{{\rm d}^2X_{\Earth}}{{\rm d}t^2}\,  \frac{\pmb{r}_\textrm{DM}}{||\pmb{r}_\textrm{DM}||} \cdot \frac{\pmb{r}_{\rm g}}{||\pmb{r}_{\rm g} ||} \right]\, ,
\end{equation}
where $g$ is the mean gravity value at the position of the gravimeter. A projection from the DM clump orbital plane to the radial direction of the gravimeter is required for the $d^2X_{\Earth}/dt^2$-term as gravimeters only measure in the radial direction.

\subsubsection{Earth Fly-by}
For close approach of a DM clump outside the Earth, the relation for the third-body acceleration becomes:
\begin{equation}\label{eq:g_r_Outside-Earth}
    \frac{\delta g_r(t)}{g} = -\frac{\mu_\textrm{DM}}{g}\left(\frac{\pmb{r}_{\rm g}-\pmb{r}_\textrm{DM}}{|| \pmb{r}_{\rm g}-\pmb{r}_\textrm{DM}||^3} + \frac{\pmb{r}_\textrm{DM}}{||\pmb{r}_\textrm{DM}||^3}\right)\cdot \frac{\pmb{r}_{\rm g}}{|| \pmb{r}_{\rm g}||}.
\end{equation}
Given this, a network of gravimeters can be simulated using the locations found in \cite{Li2020} in a similar manner as done with the Galileo constellation.

\subsubsection{Gravimeter signal}\label{subsec:gravimeter_signal_res}

\begin{figure}[h]
    \centering
    \includegraphics[width =0.45\textwidth]{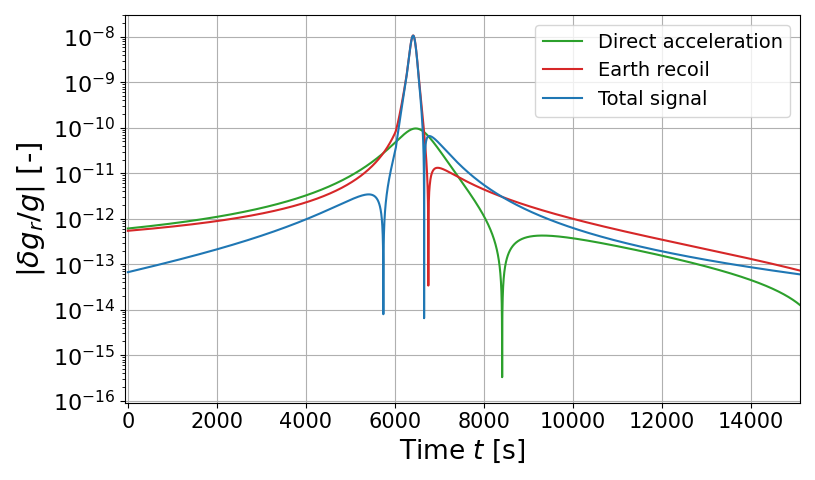}
    \vspace{-0.2cm}
    \caption{Example of a numerically simulated gravimeter total residual signal (blue line), as it would be seen at the Membach station (Belgium, $\theta_\ell = 50.609$ deg, $\Lambda_\ell = 6.010$ deg), for a clump ($10^{15}$ kg) passage inside of Earth with $B = 5000$ km, $i_{\rm DM} = 30^{\circ}$ and $V_\infty = 9$ km/s. The separate contributions of the Earth's recoil and the radial component of direct acceleration are the red and green lines respectively.}
    \label{fig:example_signal_gravimeters}
\end{figure}

Considering the coordinate ($\Lambda_\ell,\theta_\ell$) of the Membach station in Belgium (\ref{def:r_g}) and a simulated trajectory of a DM clump crossing the Earth (\ref{eq:r_DM_Earth-crossing}), an example of a gravimeter reading is shown in \autoref{fig:example_signal_gravimeters}. Unlike GNSS satellites, the nature of the signature is a transient peak.
 The Earth recoil or inertial term dominates for trajectories inside the Earth by nearly two orders of magnitude. The reason is the increase of the $m_\textrm{DM}/M_{enc}(r)$-ratio, in \autoref{eq:d2X_dt2}, as the clump penetrates the Earth. Furthermore, the local minima observed in \autoref{fig:example_signal_gravimeters} corresponds to moments when the gravimeter is blind, as described by $\pmb{r}_g\cdot \pmb{r}_\textrm{DM} \approx 0$.

\begin{figure}[h]
    \centering
    \includegraphics[width =0.42\textwidth]{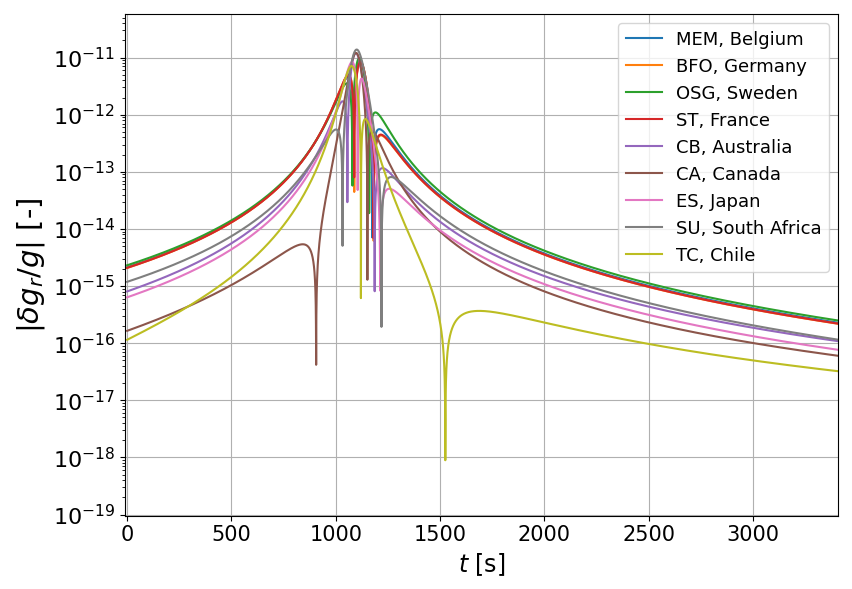}
    \caption{Gravitational perturbation reading as seen from a network of gravimeters  for a clump ($10^{15}$ kg) passage with $B=15500$ km, $V_{\infty} = 300$ km/s and $i_\textrm{DM} = 90$ deg.}
    \label{fig:network_signal_ex}
\end{figure}

Beyond the example for a single gravimeter station, the results for a worldwide network of gravimeters are visualised in \autoref{fig:network_signal_ex}. The signature of a DM clump flyby is obtained by combining the modeling of the DM clump orbit, the simulated trajectories of the diverse gravimeter stations (\ref{def:r_g}) and the radial perturbation of the gravity field induced by the DM clump (\ref{eq:g_r_Outside-Earth}).
Again, this forms the basis for observing the correlation between the signals from the different gravimeters when considering different scenarios related to the incidence direction, the closest approach distance and the excess velocity of the DM clump.

The signal shape and strength are highly sensitive to the latitude and longitude of the station. Some gravimeters would observe a double peak whereas others only a single one when the DM is at its closest distance of approach. The amplitude of the peak shows a maximum difference of nearly an order of magnitude between different stations. Hence, some gravimeters could simply not observe a significant event whereas other stations close to each other, so with expected similar signatures, e.g. BFO, Membach, OSG and ST, could be used as cross-validation or rejection of an event corresponding to a DM or another massive object flyby. A positive detection could then be further confirmed if the GNSS constellations would also provide a relevant signal.

\subsubsection{Sensitivity analysis for different parameters}\label{subsec:model_sens_res}



\begin{figure}[h]
    \centering
         \includegraphics[width=0.5\textwidth]{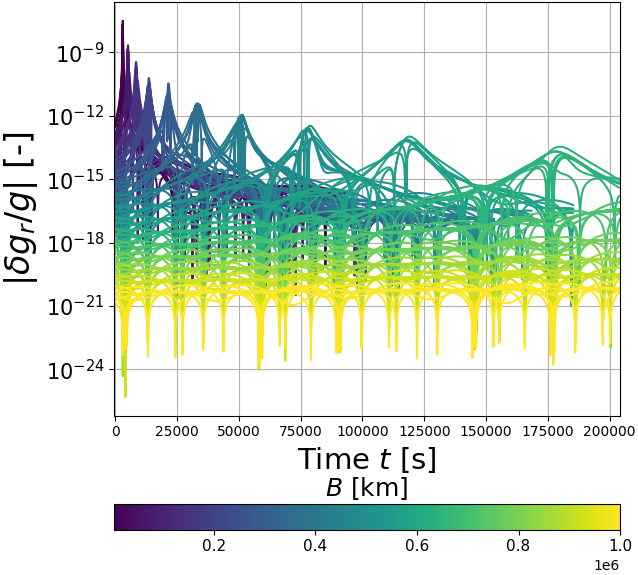}
        \caption{Entire gravimeter network signal variation as a function of $B$ with $V_{\infty}$ and $i_\textrm{DM}$ fixed at 15 km/s and 30$^\circ$.}
        \label{fig:gravi_sens_B}
\end{figure}

\begin{figure}[h]
         \centering
        \includegraphics[width=0.5\textwidth]{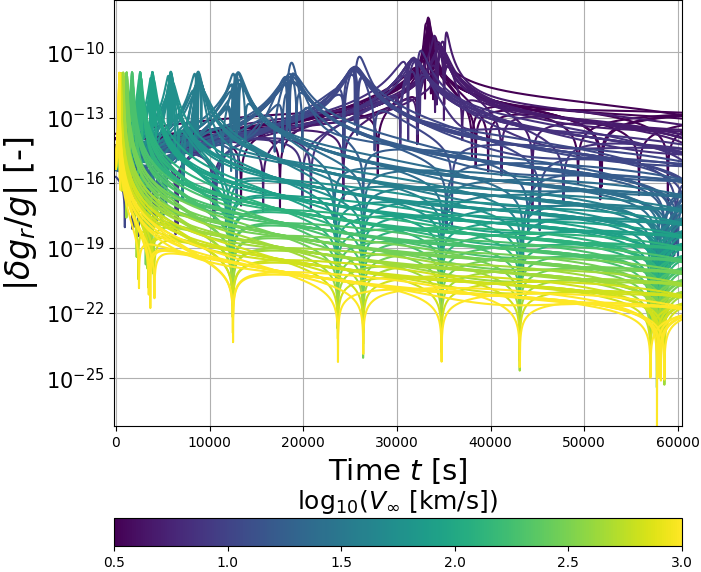}
        \caption{Entire gravimeter network signal variation as a function of $V_{\infty}$ with $B$ and $i_\textrm{DM}$ fixed at 15500km and 30$^\circ$.}
        \label{fig:gravi_sens_V_inf}
\end{figure}

In this subsection, we visualise how the radial perturbation of the gravity field $\delta g_r$ is sensitive to a given change in the DM orbital parameters, focusing on the effect of $B$ and $V_{\infty}$ in \autoref{fig:gravi_sens_B} and \autoref{fig:gravi_sens_V_inf}. This is important for future statistical data analysis since specific patterns for different gravimeter position or values of $B$ and $V_{\infty}$ would suggest a template bank study for a future statistical analysis.

As expected in case of Earth flyby, an increase in $B$ (\autoref{fig:gravi_sens_B}) causes an decrease in the amplitude of the signal due to the $1/r^2$ term in \autoref{eq:g_r_Outside-Earth}. In parallel the signal duration is approximately proportional to the value of $B$.
As far as the velocity is concerned, as $V_{\infty}$ increases, the signal's duration decreases. Obviously, the faster the clump, the shorter the signal. The amplitude is slightly increasing because slower DM clumps have a higher probability to encounter more closely gravimeters on the ground. The reason is the stronger deflection of the DM clump trajectory towards the Earth surface. 
Finally, as described in Appendix \ref{app:signature}, the expected signature is strongly dependent on the DM clump orbital orientation, due to the $\pmb{r}_g\cdot \pmb{r}_\textrm{DM}$-term. The high variability in the maximum amplitude, by one to two orders of magnitude, can limit a potential of stacking between stations but favors the orbit reconstruction in template matching analysis.

\section{Sensitivity analysis}\label{sec:Sensitivity}
In this section, a preliminary analysis based on superconducting gravimeter data series and Galileo orbital solutions will provide a first assessment of the ‘one-probe’, i.e. one satellite or one gravimeter, sensitivity. More precisely, a threshold level (per single probe) will be assumed by computing the PSD (Power Spectral Density) of the semi-major axis for satellites and gravity residuals for gravimeters as done in \cite{Rosat2003}, as a function of sampling frequency $f$. In our analysis, the PSD is computed using the \textit{scipy.signal.welch} module function in Python based on the Welch method \cite{scipy_welch, Welch1967}. From this PSD, the standard deviation, $\sigma_{\rm PSD}$, of the signal for a given frequency is computed as follow:
\begin{equation}\label{eq:sigma_PSD}
    \sigma_{\rm PSD} = \sqrt{\frac{f}{2} \, {\rm PSD}}\, .
\end{equation}
Finally, the one-probe sensitivity to gravitational disturbance induced by a DM clump flyby will be determined by the overlapping area between the measured $\sigma_{\rm PSD}$ and the area of simulated events, called the envelop, for given values of $B$, $V_\infty$ and $m_{\rm DM}$.


\subsection{Data}

\subsubsection{GNSS data}
Precise GNSS satellite orbits are computed by the International GNSS Service since 1994. The precision has dramatically improved in the first years to reach a level of stability during the last decade. 
There are different levels of precision between orbit solutions for different constellations, due to the poorer receiver coverage for more recent constellations such as Galileo and BeiDou.
GPS is currently the constellation offering the best orbit determination. GNSS satellite orbits are computed on a daily basis. In order to explore the precision and accuracy of International GNSS Service (IGS) orbits for GPS satellites, the authors in \cite{Griffiths:2009}, differentiated the geocentric positions of satellites midway between successive daily solutions of the orbit over a period of more than a year. The resulting precision ranges between 12 mm and 18 mm each direction, depending on the satellite. 
The official IGS orbit products accuracy is 25 mm for a given satellite.

The Galileo orbit solutions used in this study have been computed by the CODE analysis center \cite{Prange2020} of the International GNSS Service (IGS), from the measurements collected in a network of ground stations distributed around the world. The satellite orbital products are reported with a sampling rate $\Delta T$ of 5 min.
The CODE satellite ephemerides are given as Cartesian coordinates of the satellite positions in the ITRF (International Terrestrial Reference Frame) which is an Earth's centered Earth-fixed reference frame (ECEF).
%
%
Given a sample week of these orbital data files, the position is first transformed into an ECI (Earth-centered Inertial) frame using a transformation matrix at a given GNSS time $t_i$\footnote{This matrix is computed using SOFA software \cite{SOFA:2021-01-25} with \textit{astropy.\_erfa} \cite{Astropy}}. Then, the velocity $\pmb{V}(t_i)$ is obtained by numerically differentiating the position:
\begin{equation}
    \pmb{V}(t_i) \approx \frac{\pmb{r}_{\rm ECI}(t_{i+1}) - \pmb{r}_{\rm ECI}(t_{i})}{t_{i+1}-t_i},
\end{equation}
where $\pmb{V}(t_{i})$ and $\pmb{r}_{\rm ECI}(t_i)$ are the vectorial orbital speed and ECI-position, and $\Delta T=t_{i+1} - t_i = 5$ min, namely the interval of the CODE orbital solutions. This can be finally translated into the semi-major axis:
\begin{equation}
    \tilde{a}(t_i) = - \left( \frac{||\pmb{V}(t_i)||^2}{\mu_{\Earth}} - \frac{2}{||\pmb{r}_{\rm ECI}(t_i)||}\right)^{-1}.
\end{equation}

\begin{figure}[h]
    \centering
    \includegraphics[width=0.5\textwidth]{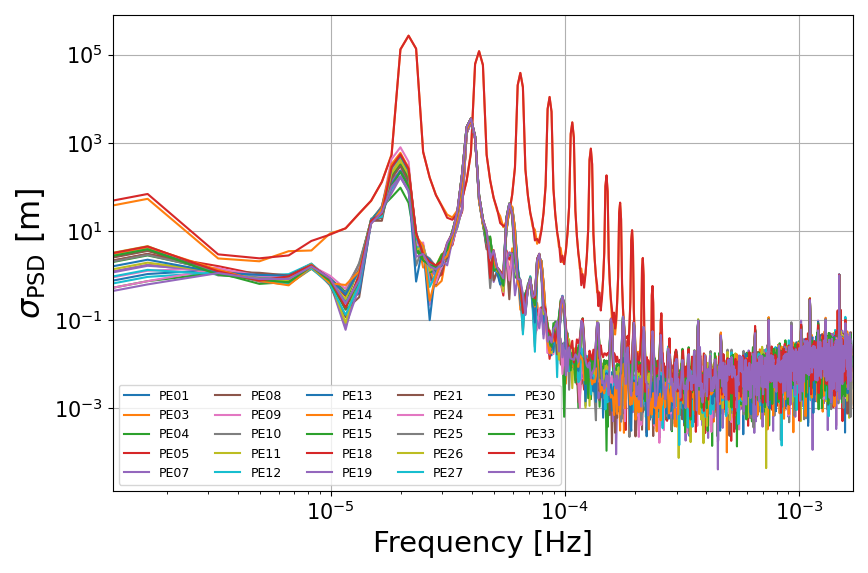}
    \caption{Frequency domain PSD of the semi-major axis time series from CODE orbital solutions (with 5 minutes interval) for a period of 7 days, from year-day 2022-2960 to 3020.}
    \label{fig:PSD_a_tilde_7days}
\end{figure}

The semi-major axis $\sigma_{\rm PSD}$ we computed from the CODE orbital solutions \cite{Prange2020} is shown in \autoref{fig:PSD_a_tilde_7days}. 
Below $10^{-4}$ Hz, the noise is dominated by systematic effects like e.g. the solar pressure radiation, with peaks corresponding to the Galileo orbital period of about 14h and its harmonics.
%
The two Galileo satellites PE14 and PE18 with eccentric orbits ($e=0.162$) experience higher periodical deviations in \autoref{fig:PSD_a_tilde_7days}. 
All satellites however have their PSDs which converge at higher frequencies with an average value of 1 cm, consistent with the current level of orbital data precision, and a mean variation raging over two orders of magnitude.

\subsubsection{Gravimeters data}

\begin{figure}[h]
         \centering
        \includegraphics[width=0.5\textwidth]{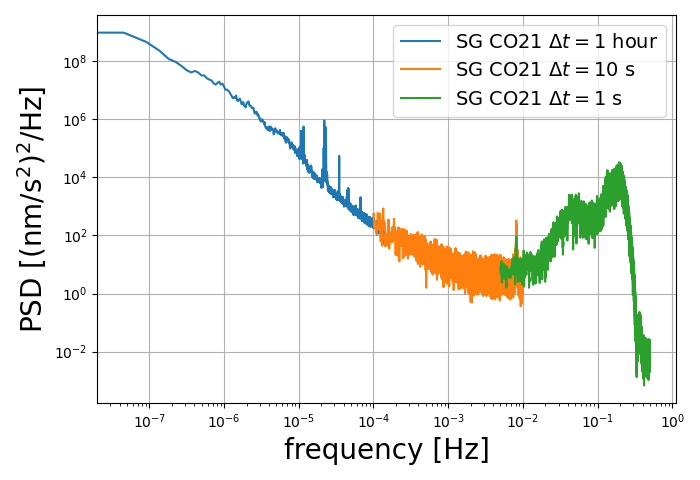}
        \caption{PSD of the time series of the Membach station superconducting gravimeter with three different sampling times used (adapted from \cite{VanCamp:2005}).}
        \label{fig:Membach_station_res_PSD}
\end{figure}

Superconducting gravimeters, outstanding broadband seismometers and spring gravimeters have a power spectral density noise level ranging typically 0.5–10 (nm/s²)²/Hz, which means that they are able to detect temporal gravity change ranging 0.06-–0.3 nm/s$^2$ 
within 1 minute \cite{VanCamp:2017-2,Ringler:2017,Rosat:2018}.
%
Peterson's New Low-Noise Model (NLNM)\cite{Peterson:1993, Berger:2004} provides the lowest noise level a seismometer or a gravimeter may achieve at the surface of the Earth. As shown in \autoref{fig:Membach_station_res_PSD}, this coloured noise experiences its lowest level at 0.5 (nm/s²)²/Hz between 1 and 10 mHz, corresponding to the instrumental noise. 
Above 10 mHz, the noise increases due to the macroseismic noise caused by the ocean swell, then decreases again above 1 Hz before reaching the instrumental limits.

\begin{figure}[h]
         \centering
        \includegraphics[width=0.49\textwidth]{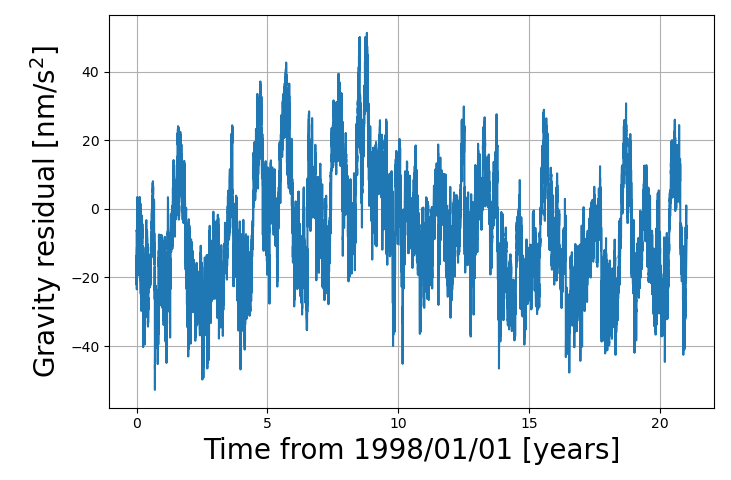}
        \caption{Gravitational residuals time series with sample time of one hour since 1998/01/01. Data from the Membach station superconducting gravimeter \cite{VC2017}.}
        \label{fig:Membach_station_res}
\end{figure}

Since 1995, the superconducting gravimeter in Membach (Belgium) measures continuously the local variations of gravity $g$ \cite{VC2017}. This will allow us to proceed to the data mining of 27 years of very high-quality, archival gravimetric data of the Royal Observatory of Belgium. \autoref{fig:Membach_station_res} shows the residuals time series for the Membach station. The residuals are obtained after correcting the time series from tidal, atmospheric and polar motion effects. 
 
\subsection{Preliminary Data Analysis}
A preliminary analysis based on the PSD of 1) the gravimeter residuals data series and 2) the semi-major axis data series of Galileo satellites provides a first assessment of the ‘one-probe’, i.e. one satellite or one gravimeter, sensitivity. For that purpose, we have modeled a physical envelope consisting in a bank template of simulated events with their characteristic amplitude, $|\delta g/g|_{\rm max}$ (\ref{eq:g_r_Outside-Earth}) and $\Delta a$ (\ref{eq:da_dg_GNSS}) for gravimeters and GNSS respectively, and duration $\Delta t_\textrm{DM}$. Then, the one-probe sensitivity is determined by the overlapping area between the PSD and the envelope of simulated events, identifying $1/f$ as $\Delta t_\textrm{DM}$.

\subsubsection{Signal duration definition}\label{subsec:signal_duration}
\begin{figure}[h]
    \centering
    \includegraphics[width=0.48\textwidth]{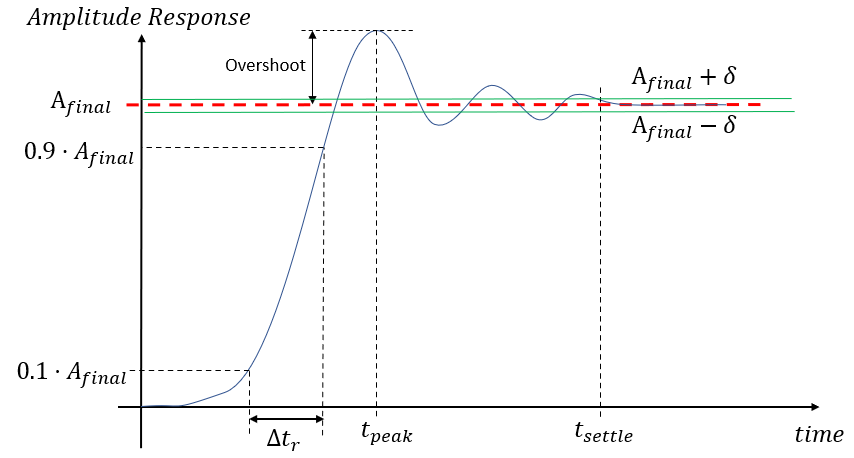}
    \caption{DM semi-major axis perturbation signal representation with specified maximum amplitude and rise and settling time, $\Delta t_r$ and $t_{settle}$ ($\delta$ is chosen to be 5\% of the final value), as estimates of the time duration of the signal.}
    \label{fig:delta_t_step}
\end{figure}

Template matching requires defining a threshold in which the signal perturbation is defined with a certain duration $\Delta t_\textrm{DM}$. For the GNSS signal, given its characteristic step-like pattern in \autoref{fig:delta_a_c_ex}, the signal duration could be defined either as the settling time or the rise time from 10\% to 90\% of the settled value, which due to the highly observed damped motion are both seen as a good indicators of the signal duration, see \autoref{fig:delta_t_step}. Both should then be analysed, by taking the PSD of one signal and checking which one is more characteristic.

\begin{figure}[h]
    \includegraphics[width=0.48\textwidth]{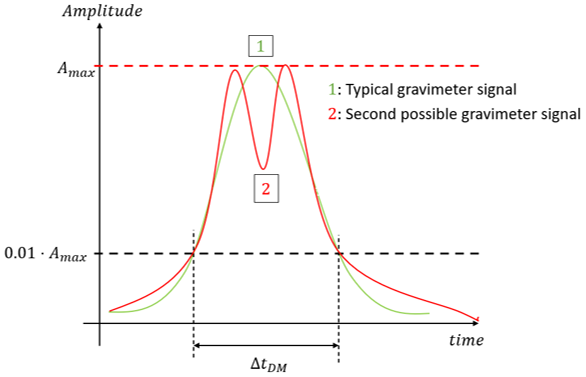}
    \caption{DM gravitational perturbation signal with specified maximum amplitude for gravimeters and definition of the time duration, $\Delta t_\textrm{DM}$, of the signal.}
    \label{fig:delta_t_DM_rep}
\end{figure}

For what concerns the gravitational perturbation induced by a DM clump on superconducting gravimeters, the $1\%$-to-maximum definition of the characteristic pulse in \autoref{fig:network_signal_ex} is taken as the duration of the signal as summarised in \autoref{fig:delta_t_DM_rep}.
%
%
%

However, this definition of the event duration could turn out to be problematic when we will proceed to data mining with genuine gravimeter data series. For a simple or double-peak signal with $\delta g/g$ amplitude of $10^{-11}$, the duration would start two orders of magnitude lower at $10^{-13}$, which is totally drowned in statistical noise. Hence, the corresponding masses and closest approach distances $d$ would not lead to a detectable signal. 
Alternatively, one can approximate the DM trajectory as a straight line which starts and ends at a distance of $10\, d$. In that case of higher velocity clumps (with $V_{\infty} > 50$ km/s) and far away from Earth: $B_{\Earth} \approx d$, leading geometrically to a lower bound: 
\begin{equation}
 \Delta t_\textrm{DM}(V_\infty,B_\Earth) \leq 2 \sqrt{99}\, \frac{B_{\Earth}}{V_{\infty}} \, ,  
\end{equation}
as $V_\textrm{DM} \geq V_{\infty}$.
This alternative definition of $\Delta t_\textrm{DM}$ is actually an equivalent estimate to the 1\% to maximum rule in \autoref{fig:delta_t_DM_rep}, due to the $1/d^2$ dependency in (\ref{eq:g_r_Outside-Earth}). Without any accessible knowledge of the bottom part of the signal below the noise, $\Delta t_\textrm{DM}$ depends on ($V_\infty$,$B_\Earth$), where $B_\Earth$ is derived from the peak amplitude, see \autoref{fig:gravi_sens_B}, while $V_\infty$ is given by the peak width, see \autoref{fig:gravi_sens_V_inf}.

\subsubsection{Physical envelop}
The physical envelop is the representation of the simulated template bank, where the events are characterised by an amplitude $|\delta g/g|_{\rm max}$ (gravimeters) or $\Delta a$ (GNSS), and a duration $\Delta t_\textrm{DM}$ as seen in \autoref{fig:envelope_gravimeters}, \autoref{fig:envelope_gravimeters2} and \autoref{fig:envelope_GNSS_rise_time}.
To simplify the data analysis, only the effect of $m_\textrm{DM}$ and $B_\Earth$ will be analysed with several fixed values for $V_{\infty}$. 
The \autoref{fig:envelope_gravimeters2} and \autoref{fig:envelope_GNSS_rise_time} illustrate the case where $V_\infty$ ranges from 2 km/s to 600 km/s. As a reminder, $V_\infty \equiv V_{\infty,\Earth}$ in this Section. Therefore, the pseudo-Maxwellian distribution of \autoref{fig:vinf_dist} is further modified by the Earth's speed of revolution around the sun of about 30 km/s, which makes lower values of $V_\infty$ more likely.
The effect of the relative angles $\omega_\textrm{DM}$ and $\Omega_\textrm{DM}$ are assumed to be included due to the different locations of gravimeter stations and Galileo satellites. 
Finally, given the standard deviation threshold computed for both gravimeter residuals and semi-major axis time series, one can compute a minimum mass $m_\textrm{DM}$, given a requirement on $B_\Earth$ (or the closest point of approach) by the overlapping area between the $\sigma_{\rm PSD}$ and the envelope of simulated events, identifying $1/f$ as $\Delta t_\textrm{DM}$. This then constitutes a first-order estimate of the current state-of-the-art sensitivity for a single probe.

\subsubsection{Gravimeter sensitivity results}\label{subsec:gravi_sens_results}

\begin{figure}[h]
    \centering
    \includegraphics[width =0.49\textwidth]{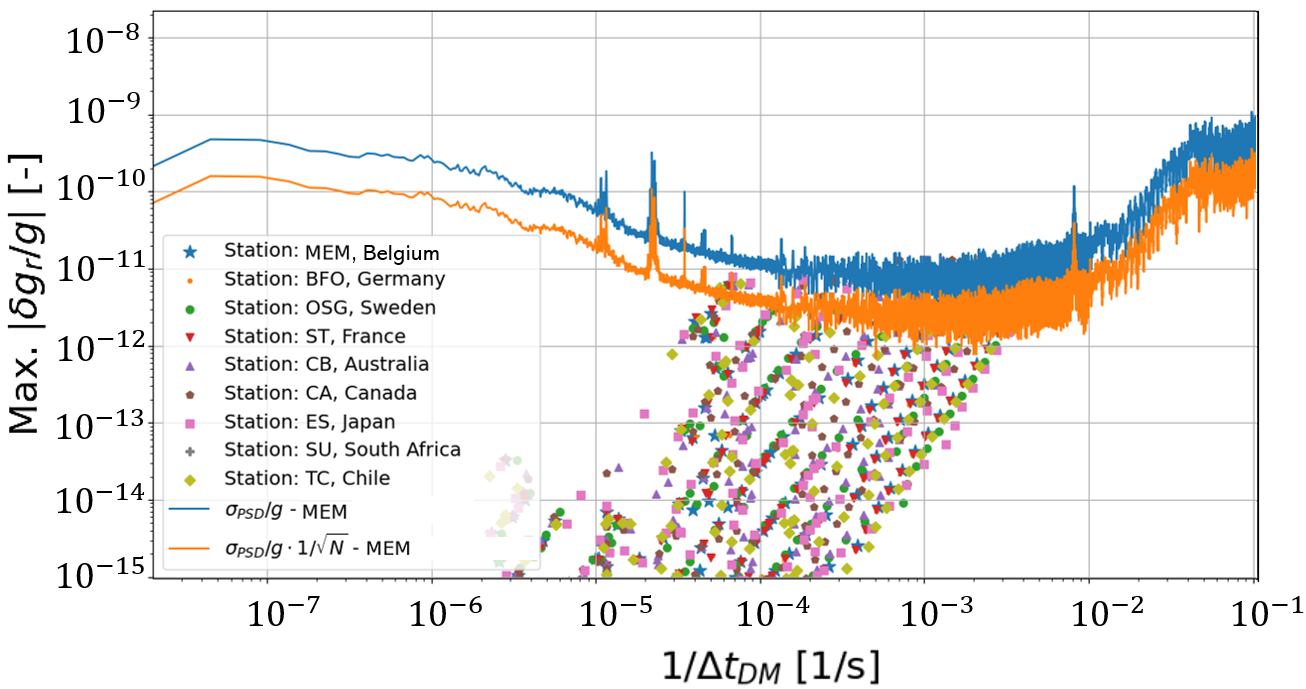}
    \caption{Physical envelope of the max. $|\delta g_{\rm r}/g|$ for a DM clump with mass $m_\textrm{DM} = 10^{15}$ kg as a function of the different stations. Added to the latter is the Membach station $\sigma_{\rm PSD}/g$-threshold level and the same threshold with an additional $1/\sqrt{N}$ correction factor}.
    \label{fig:envelope_gravimeters}
\end{figure}
\begin{figure}[h]
    \centering
    \includegraphics[width =0.5\textwidth]{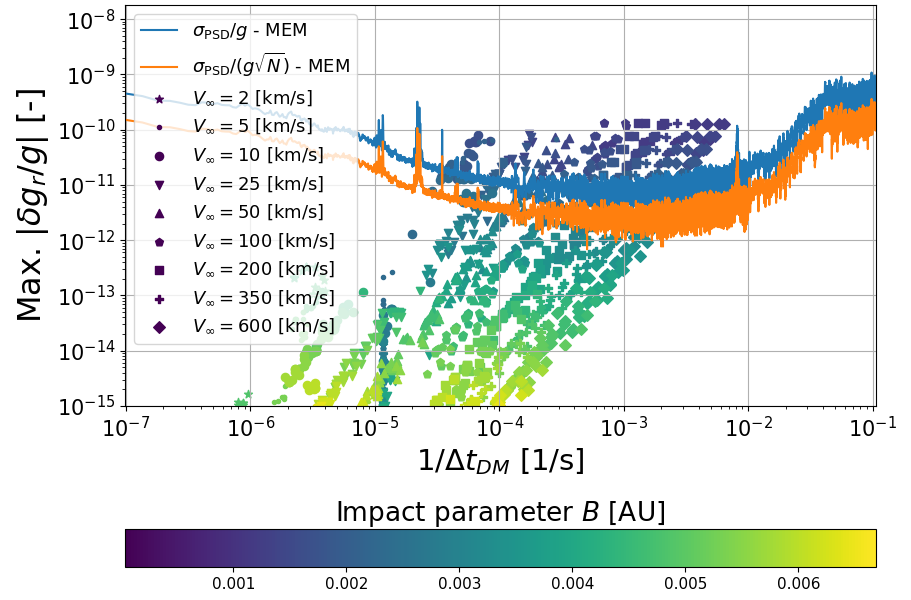}
    \caption{Physical envelope of the max. $|\delta g_r/g|$ of the clump with mass $m_\textrm{DM} = 10^{16}$ kg, for a network of 9 gravimeters, realised for 9 excess velocities (ranging from 2.5 km/s to 600 km/s) with each velocity associated to 20 impact parameters in the range [$10^{-4}$- $10^{-2}$] AU. Added to the latter is the Membach station $\sigma_{\rm PSD}/g$-threshold level and the same threshold with an additional $1/\sqrt{N}$ correction factor}
    \label{fig:envelope_gravimeters2}
\end{figure}

The resulting envelope for the $|\delta g_r/g|_{max}$ can be seen in \autoref{fig:envelope_gravimeters} and \autoref{fig:envelope_gravimeters2}, with the normalised threshold level $\sigma_{\mathrm{PSD}}/g$ for the Membach station and $i_\textrm{DM} = 23.5^{\circ}$. This specific choice was made to obtain a null heliocentric inclination relative the 23.5$^{\circ}$ tilt angle of the Earth's rotation axis. Indeed it can be shown in Appendix \ref{app:signature} that DM clumps whose orbital plane is close to the ecliptic plane are the most likely to approach the Earth at close range. There is finally the potential of stacking with a network of $N$ gravimeters \cite{Hu2020} which could lead to an additional factor of sensitivity of $1/\sqrt{N}$ to the $\sigma_{\mathrm{PSD}}$.
From both \autoref{fig:envelope_gravimeters} and \autoref{fig:envelope_gravimeters2}, we observe that the range of the signal in the frequency domain is more or less constant and frequencies of $5\cdot 10^{-5}$ to $10^{-2}$ Hz are the relevant required frequencies for detection. As a conclusion, signals with a duration between $10^2-10^4$ s have a higher potential to be differentiated from usual hydro-geological and other noise-related signals. 

Considering the diverse stations of \autoref{fig:envelope_gravimeters}, the maximum gravitational perturbation is mainly governed by:
\begin{equation}\label{eq:delta(g_r)}
    |\delta g_r|_{\rm max} \approx \frac{\mu_\textrm{DM}}{r_{\rm min}^2}\, |\text{cos}(\alpha_{r_{\rm min}})|,
\end{equation}
where $r_{\rm min}$ is the closest approach distance relative to the gravimeter station and $\alpha_{r_{\rm min}}$ is the relative angle between $\pmb{r}_{\rm g}$ and $\pmb{r}_\textrm{DM}$ at the point of closest approach. We obtains $|\delta g_r/g|_{\rm max} \approx 3.02\, 10^{-11}\, \text{cos}(\alpha_{r_{\rm min}})$ for a mass $(m_\textrm{DM})_{\rm min} = 10^{15}$ kg and $r_{\rm min} = 15000$ km, namely the highest points in \autoref{fig:envelope_gravimeters}. 
Given this, the different stations and clump velocities will allow spanning different $\alpha_{r_{\rm min}}$-values which can be seen to lower the minimum possible detection threshold. 

What's more, due to the dominance of stochastic processes caused by the hydrosphere, atmosphere or of tectonic origin on the gravitational signal (see \cite{VanCamp:2017-2}) and also taking into account a time lag between stations (see \autoref{fig:envelope_gravimeters}), only masses greater than $10^{16}$ kg can be detected at a distance $d$ of 15,000 km.

\subsubsection{GNSS sensitivity results}\label{subsec:GNSS_sens_results}


\begin{figure*}[ht]
    \centering
    \includegraphics[width =0.75\textwidth]{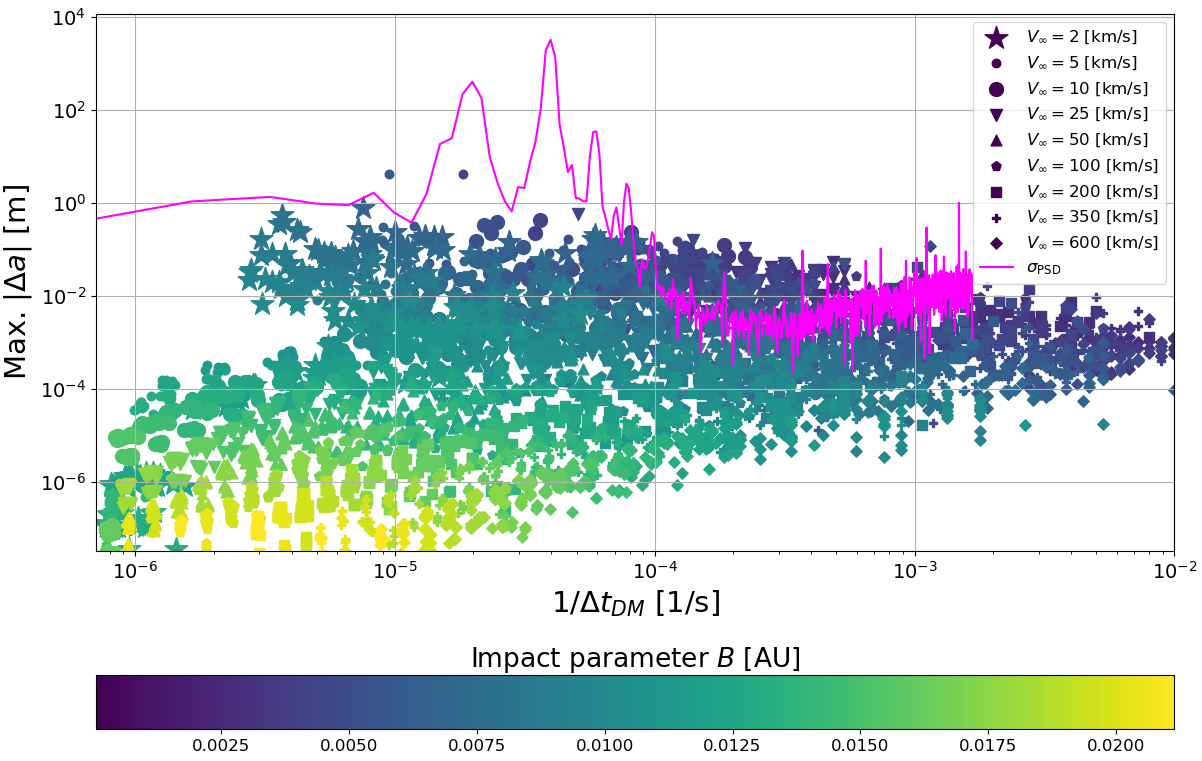}
    \caption{Physical envelope of the max. $|\Delta a|$ of the clump with mass $m_\textrm{DM} = 10^{16}$ kg, for the entire Galileo constellation, realised for 9 excess velocities (ranging from 2.5 km/s and 5 km/s to 600 km/s) with each velocity associated to 20 impact parameters in the range [$10^{-4}$- $2.1\cdot 10^{-2}$] AU. Added to the latter is the $\sigma_{\mathrm{PSD}}$-threshold level of one Galileo satellite (PE01). The rise time is the reference time of the signal.
    }
    \label{fig:envelope_GNSS_rise_time}
\end{figure*}

Due to the brief nature of the fly-by, the perturbation can be approximated as an impulse, resulting in an energetic change in orbit. It can be seen in \autoref{fig:delta_a_c_ex} that a step-like pattern is expected in the semi-major axis,  which would be highly favourable for future template matching analysis.
\autoref{fig:envelope_GNSS_rise_time} shows the resulting envelope for the $|\Delta a|_{\rm max}$ as a function of the inverse rise time, with the threshold level $\sigma_{PSD}$ for the entire constellation and $i_\textrm{DM} = 23.5^{\circ}$ (for the same reasons as introduced in the section dedicated to gravimeters).
Signals with a duration between $300-5000$ s have a higher potential to be differentiated from typical orbital perturbations and other noise-related signals, which can be seen at frequencies of $10^{-5}$ to $10^{-4}$ Hz (related to the satellite's orbital period). Indeed, the main origin of systematic effects, like e.g. the solar pressure radiation or Earth Albedo, are well modeled and already taken into account in the CODE orbit solutions. Most systematic effects are found for durations corresponding to the orbital frequency (or its first multiples). Then the sensitivity of GNSS is optimal for events of duration well below the satellite orbital period.

The \autoref{fig:envelope_GNSS_rise_time} shows that there exists a threshold $B_0(V_\infty)$ for the impact parameter $B$, depending on $V_\infty$, below which the signal amplitude becomes almost independent of $B$. This is due to the fact that the DM clump intercepts the sphere bounded by the GNSS orbits, and so only the Earth's recoil term directly decreases as $B$ increases in the equation (\ref{eq:da_dg_GNSS}).  
Following equation (\ref{eq:pericenter}), the hyperbolic pericenter $d$ is beyond the Galileo constellation's orbital radius when $B>0.0075$ AU for all $V_{\infty} > 2$ km/s. 
Conversely, for values of $V_\infty$ in the upper tail of the distribution, $V_{\infty} \rightarrow \infty$, $d \rightarrow B_\Earth$ and so the threshold $B_0 \approx 30 000$ km.

Finally, for impact parameters above the threshold $B_0(V_\infty)$, the maximum signal amplitude decrease as $B$ increases in a similar way to gravimeters.


Considering the overlapping between the PSD and the envelop of simulated events, the smallest mass that could be probed by a single Galileo satellite is about 10$^{15}$ kg for a flyby distance below 30 000 km. 

\subsection{Conclusion on sensitivity}
\begin{figure}[h]
    \centering
    \includegraphics[width=0.45\textwidth]{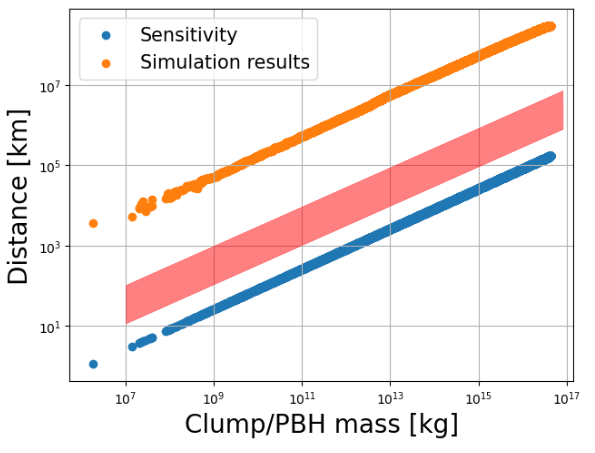}
    \caption{Orange dots: Simulation result of DM clump distance and integrated mass computed over a 20 year-period obtained from the Monte-Carlo simulation. Blue dot: Sensitivity to detection distance for a single probe as a function of the accumulated clump mass over a 20-year period. Red area: Estimation of the sensitivity enhancement using an exhaustive statistical analysis based of the correlation of the different probes}
    \label{fig:distance_sens_M}
\end{figure}

Following our fast sensitivity analysis, it turns out that individual GNSS satellites are slightly better sensors to close DM clump flyby than a gravimeter station, despite their similar intrinsic sensitivity. The reasons are the larger detection area, namely a sphere at the orbital radius, the event characteristic jump-like pattern and the lack of short-term local systematic effects. Focusing on GNSS, Fig.~\ref{fig:distance_sens_M} relates the single-satellite sensitivity (blue dots) to a minimum clump mass and a corresponding flyby distance.

In addition to the single-probe sensibility, \autoref{fig:distance_sens_M}
shows the simulation results (orange line) of the expected DM-clump / Earth encounters over a period of 20 years, given the measured density of DM in the solar system neighbourhood as a starting point. These results have been obtained from: 
\begin{enumerate}
    \item the simulation of the hyperbolic trajectories relative to the simple passage of DM clumps in the solar system, given $\rho_{\rm DM}$,
    \item the steady states density $\Delta \rho_{\rm DM}$, taking into account the captured DM clump in the solar system and the sun gravitational focusing.
\end{enumerate}
In terms of the amount of mass that can be observed at a given distance, a distance of $15,000$ km corresponds to a mass flux of $10^6$ kg/year, see \autoref{fig:m_dot_year}. This is too small to be detected at this distance on the human lifetime scale with GNSS constellation and gravimeters, since $\delta g \sim 3\cdot 10^{-19}$ m/s$^2$ for an event occurrence of 1 year. Similarly, an averaged size asteroid of $10^{10}$ kg would be observed at a distance of $10^{-2}$ AU with a signal amplitude of $ \sim 10^{-19}$ m/s$^2$.

However, this single probe analysis paves the way for a future statistical analysis based on correlations between a constellation of probes using the 28-years of publicly available GNSS and gravimeter data. It turns out that GNSS slightly outperform gravimeters due to the characteristic step-like signal against the transient peak signal shape in gravimeters, and thus providing a clear pattern for future data analysis. Such a statistical analysis would enable to gain between 2 and 3 orders of magnitude in sensitivity, see the Purple area in~\autoref{fig:distance_sens_M}).

\begin{figure}[h]
    \centering
    \includegraphics[width=0.45\textwidth]{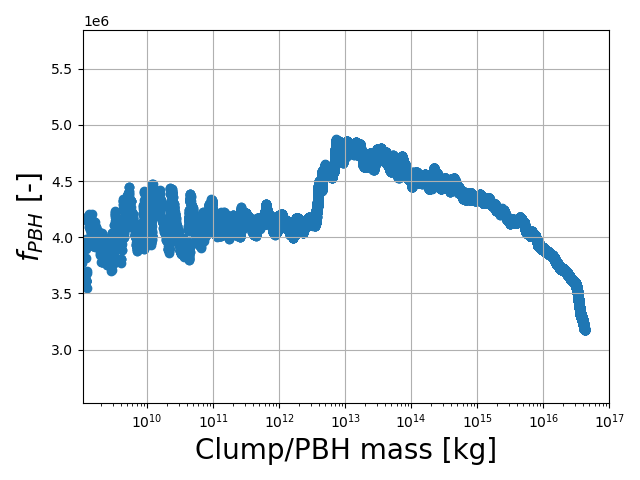}
    \caption{Forecast of the detectable dark matter fraction made of PBHs $f_{\rm PBH}$ for $20$ years of GNSS data, based on the one-probe sensitivity.  This illustrates the required improvement of the sensitivity to lead to significant constraints with $f_{\rm PBH} < 1$.  }
    \label{fig:fPBH}
\end{figure}

Despite the gain of an exhaustive statistical analysis, the ratio with our simulated event rates (orange dots in~\autoref{fig:distance_sens_M}) is around 3-4 orders of magnitude, so still not enough to hope a direct detection, considering the measured DM density $\rho_{\rm DM}$ in the solar system neighbourhood, as confirmed by \autoref{fig:fPBH}. However, such bounds on the DM clump density in the solar system would be a premiere and the best to be obtained by direct observation on a terrestrial scale for the mass range $10^{11}$-$10^{17}$ kg, before LISA is operational~\cite{Baum:2022duc}.

\section{Conclusions and perspectives} \label{sec:Conclusion}

This paper presents an exploratory study of the occurrence frequency of near-Earth trajectories of asteroid-mass DM clumps or PBHs.  Their expected velocity and minimum distance distributions have been calculated in the case of a simple passage across the solar system (SS) and for strongly elliptical orbits around the Sun after a three-body gravitational capture. Our analysis also includes an estimation of the ejection rate and the resulting steady state mass in the SS.  We have shown that one does not expect a significant increase of the number of near-Earth trajectories for captured objects.  

In the second part of the paper, we analyzed two possible detection methods of such objects and modeled how the tiny changes in the gravitational field should alter the orbit of GNSS satellites or could be detectable in the time correlations of a worldwide network of superconducting gravimeters.


The modelling of DM clump or PBH trajectories in the SS with Monte-Carlo simulations and semi-analytical methods, detailed in Sections \ref{sec:Trajectories} and \ref{sec:DM_capture}, led us to estimate the occurrence of near-Earth trajectories and the resulting potentially detectable signatures in GNSS or gravimeter networks.  Inversely, the sensitivity of GNSS constellations and gravimeters has been used in combination with these methods to forecast constraints on the relative dark matter abundance of clumps and PBHs as a function of their mass, in absence of detection.  Moreover, we assessed for the first time the expected signal-to-noise ratio based on the PSD of gravimeter residuals and on Galileo orbital solutions. The single probe (i.e. one satellite or one gravimeter) sensitivity has been calculated in relation with the fly-by characteristics. However, for having a number of detectable events of order one per year, one needs a galactic density millions times larger than that of DM, even after considering the gravitational-focusing effect by the Sun. 

%
Nevertheless, our work paves the road for an exhaustive statistical analysis of existing GNSS and gravimeter data, combining stacking, template match-filtering techniques and time-correlations between satellites or gravimeters, as well as cross-correlations between probes.  Such techniques could importantly improve the network sensitivity, possibly by two orders of magnitude compared to the single-probe one. Their efficiency would also rely on the ability to disentangle systematic effects and various noise sources. 
Although still below the threshold of detectability based on local galactic density measurements, such a result would still be remarkable and unprecedented.
It can be translated in a detection capability in scenarios where the total DM clump or PBH mass is about $8 \times 10^{18}$ kg within a sphere of 3 AU radius, for individual clump masses between $10^{8}$ and $10^{17}$ kg.  For comparison, this represents a mass four hundred times less than that of the main asteroid belt orbiting within the same radius.

GNSS networks seem the most promising for observing near-Earth DM clumps or PBHs, because these would induce a typical step-like change in the semi-major axis of satellite orbits that would be observed in the entire constellation.  By monitoring the amplitude and duration of this altitude variation for each satellite, one could in principle reconstruct the mass and trajectory of the object, an information that could be used in combination with electromagnetic observations to distinguish a near-Earth dark object from a small asteroid.
Even if the present technology does not allow to set stringent limits on DM models, unique constraints could be established on the abundance of dark objects in the SS.  These constraints would be complementary to the ones from planet ephemeris\cite{Tran:2023jci} whose range of validity is limited to DM clumps of mass larger than $10^{16}$ kg.  Finally, the simulated trajectories and population properties, the analysis techniques developed for GNSS or gravimeter networks, could also be applied to the future Laser Interferometer Space Antenna (LISA)~\cite{Baum:2022duc}.  The sensitivity of LISA to changes in the gravitational field at frequencies of order $10^{-4}$ Hz will be unprecedented and should allow to bring stringent constraints on the density of dark matter clumps and PBHs.


Our analysis could be improved or extended in several ways.  First, we only considered the case of point-like objects but our calculations could be generalized to extended dark structures of size similar or larger than the Earth. Interesting constraints could also be established if the dynamics is influenced by a fifth-force interaction of Yukawa-type. By analogy with the simulations of \cite{Hall:2016usm, Baum:2022duc} for $\mu$Hz GW detectors, it should be possible to detect or constrain DM clumps with 30 years of GNSS and gravimeter data for a fifth-force between matter and the dark sector about $10^3$ times stronger than gravity. Further investigation would be needed to calculate the effects on the trajectories in the Solar System (SS).  Enhanced gravitational focusing and a new equilibrium in the capture vs. ejection process in the SS would boost the occurrence of near-Earth trajectories. Some exotic dark matter scenarios could also lead to a local $10^6$-enhancement of the density, leading to observable signatures for gravimeters and GNSS.  
%

The Peterson's New Low Noise Model (NLNM) \cite{Peterson:1993, Berger:2004}, used for our estimations of the gravimeter sensitivity, is based on the selection of data segments during specifically quiet periods. Earthquakes, anthropogenic perturbations, storm surges, rainfall, and atmospheric pressure changes, are all causes of increased noise at levels above the NLNM. 
All these perturbations and systematic effects do not exist on the Moon.
Apart from the fall of meteorites, a well-thermally-insulated instrument would work in a much quieter environment, and not only in the 1-10 mHz frequency band but at higher and lower frequencies. Hence, the lunar deployment of broadband seismometers or gravimeters would allow stacking measurements and lower the instrumental white noise effects.  The subsequent flattening and lowering of the $\sigma_{\rm PSD}$ curve in \autoref{fig:envelope_gravimeters} and \autoref{fig:envelope_gravimeters2} at higher frequency values would improve the sensitivity to lower DM clump masses, below $10^{15}$ kg.

Finally, we point out that similar analysis could be performed for satellites equipped for Satellite Laser Ranging (SLR) \cite{Pearlman2019}, such as GRACE (Gravity Recovery And Climate Experiment), JASON, LAGEOS or Sentinel.  For instance, GRACE and GRACE-FO (GRACE Follow-On), are two-satellite systems with two different types of on-board sensors: a Ka-Band (microwave) antenna and a laser ranging interferometer (LRI). Tiny changes in the gravitational field of order  nm/s$^2$ \cite{Ghobardi_Far2020, Han2021c}, induced by ground water storage changes, glacial isostatic adjustments and earthquakes, can result in detectable range-rate variations between the two satellites. Such a technique could also probe near-Earth trajectories of DM clumps or PBHs and would be complementary to gravimeters and GNSS. 

Overall, the most fascinating -- even if still futuristic -- perspective of our work is to uncover a range of possibilities for in-situ studies of black holes or dark matter objects in the Earth vicinity, which would have groundbreaking implications for various disciplines of Physics. 


\begin{acknowledgments}
The authors thank C. Ringeval and O. Minazzoli for their time spent in very useful discussions.   The work of S.C. is supported by the \textit{Fond National de la Recherche Scientifique} (FNRS-F.R.S.) with a \textit{Mandat d'impulsion scientifique} (MIS) and by the Belgian Francqui Foundation with a Francqui Start-up Grant.
\end{acknowledgments}

\bibliographystyle{apsrev4-1}

\bibliography{biblio.bib} 

\pagebreak

\appendix
\newpage

\section{Monte-Carlo simulations in the case of a single passage} \label{app:Monte-Carlo}
\subsection{Input and output variables}
The set of used variables for the Monte-Carlo are summarised in \autoref{tab:variables_1}, where for Earth, $\Omega$ and $\omega$ are not needed due to the assumed circular nature of the orbit. The desired outputs are also shown. 
%
\begin{table*}[ht]
\centering
\caption{Keplerian elements and additional variables related to the DM clump and Earth orbits, with final desired outputs for a total population size of $2\cdot 10^6$.}
\label{tab:variables_1}
\begin{tabular}{m{6.2cm}m{4.8cm}m{1.9cm}}
\hline
\textbf{Orbital variables} & \textbf{DM clump orbit}       & \textbf{Earth orbit}  \\ \hline
Semi-major axis, $a$ & Via $V_{\infty}$, see \autoref{fig:vinf_dist} & $a_{\Earth} = 1$ AU          \\ 
Inclination, $i$ [rad] & $\mathcal{U}[0,\pi]$ and $\mathcal{U}[0,8.72\cdot 10^{-4}]$ & $i_{\Earth} = 0$   \\ 
RAAN, $\Omega$ [rad] & $\mathcal{U}[0,2\pi]$ & N.A. \\ 
Argument of periapsis, $\omega$ [rad] & $\mathcal{U}[0,2\pi]$ & N.A.\\ 
Initial true anomaly, $\theta_0$ [rad] & $\theta_0 = \pm 0.97\, \theta_{\rm lim}$  & $\mathcal{U}[0,2\pi]$   \\ 
Impact parameter, $B$ [AU] & $\mathcal{U}[0.01, 100]$ and $\mathcal{U}[0.98, 1.025]$ & N.A.  \\ 
\hline
\textbf{Desired outputs}  & \textbf{Variable symbol}    &  \textbf{Unit}  \\ 
\hline
Minimum Earth/DM clump distance & $d$ & [AU]  \\ 
DM velocity (w.r.t Earth) at distance $d$ & $V_{DM/\Earth}$ & [km/s]  \\ 
\hline
\end{tabular}
\end{table*}

In addition, the relative velocity $V_{\textrm{DM}/ \Earth}$ is simply computed at the specific time of closest approach, see (\ref{def:VDM/E}). It must be noted that for this problem, a variable time step is used due to the difference between the time scales of the Earth and DM clump orbits. This time step reads: 
\begin{equation}
    \Delta t(t) = \epsilon \cdot  \frac{r_{\textrm{DM}/\Earth}(t)}{V_{\textrm{DM}/\Earth}(t)}, 
\end{equation}
with an initial condition set to $\Delta t_0 = \epsilon$ and $\epsilon=0.03$, chosen for convergence of the Runge-Kutta 4 integrator.

Finally, the initial true anomaly is set to be: $\theta_0 = -q\, \theta_{\rm lim}$ with $q$ a numerical parameter set to $q = 0.97$ to avoid the singularity of \autoref{eq:keplerian_radius}, but still sufficient to simulate efficiently a high initial $r_{\textrm{DM}}(t_0)$.

\subsection{Influence of the inclination angle}
\begin{figure}[h]
    \centering
    \centering
    \includegraphics[width=0.45\textwidth]{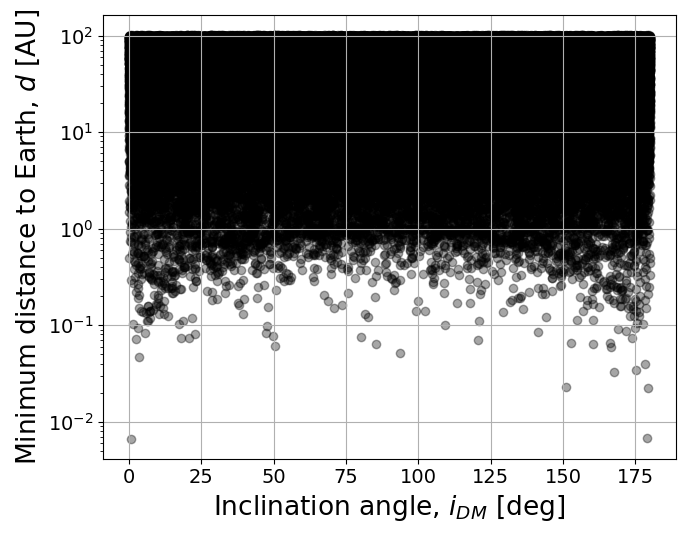}
    \caption{Resulting populations from four independent Monte-Carlo simulation of minimum Earth-Clump distance as a function of $i_\textrm{DM}$ for the general distribution of impact parameters.}
    \label{fig:d_i_DM_s}
\end{figure}

The influence of the inclination angle is shown in \autoref{fig:d_i_DM_s}.
The smallest distances are obtained for velocities within the range of $100-500$ km/s (\autoref{fig:m_dot_year}) and $i_\textrm{DM} \sim 0^\circ$ or $180^\circ$ (\autoref{fig:d_i_DM_s}).

As expected, while approaching $i_{DM} \rightarrow 0$ or 180 deg, the minimum distance to Earth can reach relatively smaller values down to $10^{-2}$ AU. Interestingly, however this effect is much smaller than anticipated as the distribution of distances is mostly uniform even at the extreme values. The effect of the inclination is therefore found to be negligible. 

\section{Calculation of the capture rate and properties of the newly bound orbit}
\label{app:Capture}
For Monte-Carlo simulation of the capture process, the variables are summarised in \autoref{tab:variables_2}.
\begin{table*}[ht]
\centering
\caption{Most important variables and outputs related to the DM clump and planet orbits for a total population size of $5\cdot 10^5$ per planet.}
\label{tab:variables_2}
\begin{tabular}{m{7.0cm}m{6.0cm}m{0.9cm}}
\hline
\textbf{Variables} & \textbf{Value or distribution}       & \textbf{Unit}  \\ \hline
Excess heliocentric velocity, $V_{\infty}$ & Distribution in \autoref{fig:vinf_dist} & [km/s]         \\ 
Impact parameter, $B$ & $\mathcal{U}[0.01B_{max}, B_{max}]$ & [AU]  \\ 
Inclination, $i_\textrm{DM}$ & $\mathcal{U}[0,\pi]$ & [rad]   \\ 
RAAN, $\Omega_\textrm{DM}$  & $\mathcal{U}[0,2\pi]$ & [rad] \\ 
Argument of periapsis, $\omega_\textrm{DM}$  & $\mathcal{U}[0,2\pi]$ & [rad]\\ 
Planet true anomaly, $\theta_P$ & $\mathcal{U}[0,2\pi]$  & [rad]  \\ 
\hline
\textbf{Desired outputs}  & \centering \textbf{Variable symbol}    &  \textbf{Unit}  \\ 
\hline
Bounded semi-major axis distribution & \centering $a_b$ & [AU]  \\ 
Bounded eccentricity distribution & \centering $e_b$ & [-]  \\ 
Probability of capture & \centering $P_c$ & [-]  \\ 
\hline
\end{tabular}
\end{table*}
%
%
%
%
Each planet contributes to the capture by a fraction:
$M_{\rm c} \approx M_\textrm{DM}\cdot P_{\rm c}$, where $P_{\rm c}$ results from the occurrence probability of captured events $P_{\rm MC,c}$ obtained from the Monte-Carlo simulations, see (\ref{def:P_c}).\autoref{tab:masses_results} shows the capture probabilities $P_{\rm c}$ and mass estimates $M_{\rm c}$ computed in this study in comparison with results obtained previously in the Literature: 
\begin{itemize}
    \item The ratio between Jupiter and Saturn of the total mass is about 1 in agreement with \cite{Xu2008} but diverging from \cite{Khriplovich2009}, where this ratio is smaller than four. In the analysis of \cite{Khriplovich2009}, Jupiter is by far the main contributor to the capture process, which is not the case in our study. 
    \item In this study, the estimate of the mass captured $M_{\rm c}$ by the Earth, Saturn, Uranus and Neptune is greatly enhancement compared with the results obtained previously in \cite{Khriplovich2009}, from one to five orders of magnitude. For example, the contribution of Uranus is no longer negligible.
\end{itemize}
With regards to \cite{Khriplovich2009}, the results for the probability capture $P_{\rm c}$ and the mass captured $M_{\rm c}$ computed in this study and summarized in \autoref{tab:masses_results} seem to be less conservative for all planets, except for Jupiter. Unlike our method, the methodology of \cite{Khriplovich2009} is entirely analytical, with mass estimates derived from analytical and probabilistic laws. The observed discrepancies will deserve further investigation.

According to \autoref{tab:masses_results}, \cite{Xu2008} shows a clear overestimation of $M_{\rm c}$, not only confirmed by the results of \cite{Khriplovich2009} and of this study, but also by the analysis of \cite{Edsjo2010}. An explanation found for this discrepancy with \cite{Xu2008} could be the capture radius not taken as the SOI, but a factor of Hill's radius as $r_{\rm H} = a_{\rm P} (m_{\rm P}/M_{\Sun})^{1/3}$. This is then associated to no clear correction by a probability of orbital intersection and finally an assumed inclination of $0$ (the latter was found to increase the probabilities by a factor of 10, except for Jupiter with an increase factor of 1.1).


Let us now briefly mention the results of our calculation of the elliptical orbits of the captured objects. 
From the analytical method, higher velocity planets provide more data points. 
The semi-major axis $a_{\rm b}$ of the newly elliptical orbit is increasing with the distance of the planet to the Sun. In some case, the extreme values $a_{\rm b} > 1000$ AU are reached. 
The newly bound orbit can also be characterised by its eccentricity $e_{\rm b}$. The eccentricity can be computed using \autoref{eq:ecc_hyp}, where the orbital energy $\mathcal{E} = - \mu_{\Sun}/(2a_{\rm b})$ and $h \approx a_{\rm P} \, (V_{\rm f})_\text{t}$, where $(V_{\rm f})_\text{t}$ corresponds to the tangential component of $V_{\text{f}}$ and $a_{\rm P}$ is the planet's orbital radius. This results in:
 \begin{equation}
     e_{\rm b} = \sqrt{1 - \frac{a_{\rm P}^2\,  (V_{\rm f})_\text{t}^2}{\mu_{\Sun} a_{\rm b}}}
 \end{equation}
 As the final velocities are usually very close to $V_{\text{esc}}$, the fractional term $a_{\rm P}^2\, (V_{\rm f})_\text{t}^2/(\mu_{\Sun} a_{\rm b})$ is very small. This leads to $e_{\rm b}$-values close to $1$.
Finally, due to the large $e_{\rm b}$ values, pericenter values are actually concentrated at low values with a maximum of only 4 AU, below Jupiter's orbital radius. This enhances the probability of their possible encounter with the Earth but also their probability of being ejected by Jupiter. 



\section{Effect of DM orbit inclination}\label{app:signature}
\begin{figure}[h]
         \centering
        \includegraphics[width=0.5\textwidth]{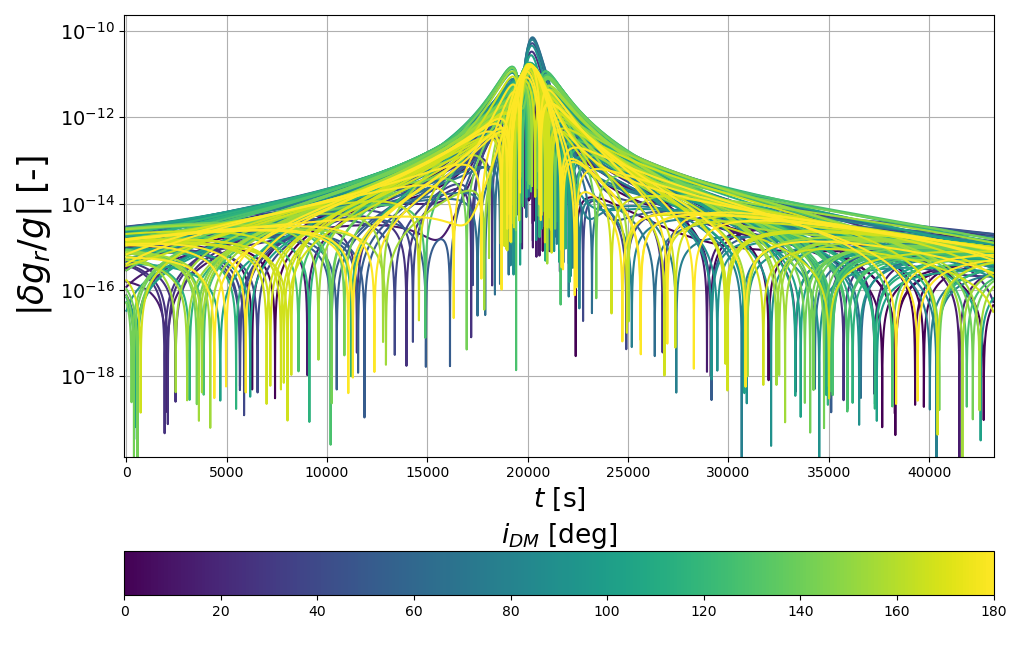}
        \caption{Entire gravimeter network signal variation as a function of $i_\textrm{DM}$ with $V_{\infty}$ and $B$ fixed at 15 km/s and 15500 km respectively.}
        \label{fig:gravi_sens_i_DM}
\end{figure}
 The change in inclination of the DM orbit in \autoref{fig:gravi_sens_i_DM} shows a large variability in the maximum amplitude of the gravimeter signal, of one to two orders of magnitude, due to the angle $\alpha_{r_{\rm min}}$ in \autoref{eq:delta(g_r)}. The stations can also be observed to have different signal shapes. The duration is however nearly unaltered (maximum factor difference of 1.5).



\end{document}